\documentclass[universe,article,accept,pdftex,moreauthors]{Definitions/mdpi}

\firstpage{1}
\pubvolume{0}
\issuenum{0}
\articlenumber{0}
\pubyear{2023}
\copyrightyear{2023}
\datereceived{16 October 2023}
\daterevised{02 November 2023} 
\dateaccepted{07 November 2023}
\datepublished{in press}
\hreflink{https://doi.org/10.3390/universe00000} 

\pdfoutput=1

\Title{Dark Matter in Fractional Gravity III: Dwarf Galaxies Kinematics}

\TitleCitation{Dark Matter in Fractional Gravity III: Dwarf Galaxies Kinematics}

\Author{Francesco Benetti$^{1,2}$*\orcidA{}, Andrea Lapi $^{1,2,3,4}$\orcidB{}, Giovanni Gandolfi $^{1,2,3}$\orcidC{}, Minahil Adil Butt$^{1,2,3}$\orcidD{}, Yacer Boumechta$^{1,2,3,5}$\orcidE{}, Balakrishna S. Haridasu$^{1,2,3}$\orcidF{}, Carlo Baccigalupi$^{1,2,3}$\orcidG{}}

\AuthorCitation{Benetti, F.; Lapi, A.; Gandolfi, G; et al.}

\address{$^{1}$ \quad Scuola Internazionale Superiore Studi Avanzati (SISSA), Physics Area, Via Bonomea 265, 34136 Trieste, Italy; fbenetti@sissa.it (F.B.); lapi@sissa.it (A.L.); ggandolf@sissa.it (G.G.); mbutt@sissa.it (M.A.B.); yboumech@sissa.it (Y.B.); sandeep.haridasu@sissa.it (B.S.H.); bacci@sissa.it (C.B.)\\
$^{2}$ \quad Institute for Fundamental Physics of the Universe (IFPU), Via Beirut 2, 34014 Trieste, Italy\\
$^{3}$ \quad Istituto Nazionale Fisica Nucleare (INFN), Sezione di Trieste, Via Valerio 2, 34127 Trieste,  Italy\\
$^{4}$ \quad Istituto di Radio-Astronomia (IRA-INAF), Via Gobetti 101, 40129 Bologna, Italy\\
$^{5}$ \quad ICTP-The Abdus Salam International Centre for Theoretical Physics, Strada Costiera 11, 34151 Trieste, Italy}

\corres{Correspondence: fbenetti@sissa.it}

\abstract{Recently we put forward a framework where the dark matter (DM) component within virialized halos is subject to a non-local interaction originated by fractional gravity (FG) effects. In previous works we demonstrated that such a framework can substantially alleviate the small-scale issues of the standard $\Lambda$CDM paradigm, without altering the DM mass profile predicted by $N-$body simulations, and retaining its successes on large cosmological scales. In this paper we dig deeper to probe FG via high-quality data of individual dwarf galaxies, by exploiting the rotation velocity profiles inferred from stellar and gas kinematic measurements in $8$ dwarf irregulars, and the projected velocity dispersion profiles inferred from the observed dynamics of stellar tracers in $7$ dwarf spheroidals and in the ultra-diffuse galaxy DragonFly 44. We find that FG can reproduce extremely well the rotation and dispersion curves of the analysed galaxies, performing in most instances significantly better than the standard Newtonian setup.} 


\begin{document}

\section{Introduction}\label{sec|Intro}

The standard $\Lambda$CDM cosmology envisages galaxies to be hosted in virialized halos of dark matter (DM), which largely dominate the total mass and hence mostly determine the overall gravitational potential well and the dynamical properties of the baryons \cite{Rubin80,Persic96}. Remarkably, the density distribution of such halos is predicted from $N-$body simulations to follow an approximately universal shape, well described by the classic Navarro-Frenk-White \cite{Navarro97} profile $\rho\propto (r/r_s)^{-1}\,(1+r/r_s)^{-2}$, with $r_s$ being a characteristic scale radius where the logarithmic slope equals $-2$. 

Only a minor deviation from such a scale-invariant behavior is expected, which amounts to a relationship between $r_s$ and the total DM mass \cite{Dutton14}. This is often expressed in terms of the concentration parameter $c_{200}\equiv R_{200}/r_s$, with $R_{200}$ being the radius where the average DM density is $200$ times that of a critical Universe $\rho_{\rm crit}$. In fact, recent zoom-in $N-$body simulations \cite{Wang20} have demonstrated that $c_{200}$ correlates very well with the mass $M_{200}\equiv (4\pi/3)\,200\, \rho_{\rm crit}\, R_{200}^3$ over the astonishingly extended range from $M_{200}\sim 10^{-5}\, M_\odot$ to $10^{15}\, M_\odot$.

Although on large scales observational data undoubtedly confirm the above picture, in the realm of dwarf galaxies with total masses $\lesssim 10^{11}\, M_\odot$ the situation becomes more uncertain. The most relevant issue for the present context emerges from galaxy kinematics and/or gravitational lensing data, which seem to indicate a much flatter density profile in the inner regions (i.e., a core) with respect to the cuspy NFW behavior; this occurrence is often referred to as the cusp-core problem \cite{Flores94,Gentile04,deBlok08,Oh15}.

In addition, there are other well-known issues associated with small galaxy scales that are worth mentioning \cite{Bullock17}: the missing satellites problem \cite{Klypin99,Moore99} concerns the observed satellites in Milky-Way sized galaxies, that are found to be much less numerous than the bound DM halos in $N-$body simulations; the too-big-to-fail problem \cite{BoylanKolchin12} concerns the halos hosting dwarf galaxies, which from kinematical measurements are found to be less massive than expected; the radial acceleration relation \cite{McGaugh16,
Lelli17}, the universal core
surface density \cite{Donato09}, and the
core radius vs. the disk scale length scaling \cite{Donato04} all constitute tight empirical relationships between the properties of the DM and of the baryons that are extremely puzzling in the standard paradigm.

There are various viable solutions to these issues. The most obvious claims a misinterpretation of the data due to poor resolution effects or other complex features in the DM distribution \cite{Oman15}. Another one invokes the impact of ordinary matter physics on the DM profile via stellar feedback \cite{Pontzen14,Freundlich20} or transfer of energy/angular momentum to the DM via dynamical friction \cite{ElZant01,Tonini06}. Another possibility involves nonstandard particle candidates such as warm or sterile neutrino DM \cite{Bode01,Lovell14}, fuzzy or particle-wave DM \cite{Hu00,Hui17}, self-interacting DM \cite{Vogelsberger16}, dark-photon DM \cite{McDermott20,Bolton22}, that by various processes (e.g., free streaming, quantum pressure effects, and/or dark sector interactions) can avoid the formation or later erase the inner cusp \cite{Bertone04,Salucci21}. Finally, the observed galaxy kinematics can be explained with or without DM by modified gravity theories \cite{Clifton12,Nojiri17,Saridakis21} such as MOND \cite{Milgrom83,Famaey12}, fractional-dimensional gravity \cite{Giusti20a,Giusti20b,Varieschi20,Varieschi21,Calcagni22},
emergent entropic gravity \cite{Verlinde17,Yoon23}.

Recently, in \cite{Benetti23a,Benetti23b} we put forward a fractional gravity (FG) framework that strikes an intermediate course between a modified gravity theory and an exotic DM scenario 
(in this respect similar to the dynamical non-minimally coupled DM model explored by our team in \cite{Gandolfi21,Gandolfi22,Gandolfi23}). FG envisages the DM component to be present though subject to a non-local interaction mediated by gravity. Specifically, in such a framework the gravitational potential associated to a given DM density distribution (e.g., the NFW one) is determined by a modified Poisson equation including fractional derivatives (i.e., derivatives of noninteger type), that are aimed at describing non-locality. Very interestingly, it can be shown that FG can be reformulated in terms of the standard Poisson equation, but with an effective density distribution which is flatter in the inner region with respect to the true one. This is actually the density behavior that an observer would infer by looking at the kinematic data and interpreting them in terms of standard Newtonian theory. Thus in FG the cusp-core problem is basically solved at its root, since the cuspy NFW density profile of $\Lambda$CDM originates in FG a dynamics very similar to a cored profile in the standard Newtonian setting.

In \cite{Benetti23a,Benetti23b} we tested FG over an extended mass range $M_{200}\sim 10^9-10^{15}\, M_\odot$ by exploiting stacked rotation curves of spiral galaxies and joint X-ray/Sunyaev-Zel'dovich observations of galaxy clusters. We found that FG performs extremely well in reproducing the data in all these systems. Moreover, our analysis highlighted that the strengths of FG effects tend to weaken toward more massive systems, so implying that FG can substantially alleviate the small-scale issues of the standard $\Lambda$CDM paradigm, while retaining its successes on large cosmological scales.

In this paper we aim at digging deeper into the regime where FG effects are expected to be more relevant, focusing on individual dwarf galaxies. In these objects the cusp-core problem is observationally very pressing, and its solution via baryonic effects is difficult to be envisaged given the paucity of baryons. Specifically, we probe FG both in irregular dwarf (dwIrr) galaxies by exploiting the rotation velocity profile inferred from stellar and gas kinematical measurements, and in dispersion-dominated dwarf spheroidals (dwSph) by exploiting the velocity dispersion profile inferred from the dynamics of stellar tracers. 

The structure of the paper is the following: in Section \ref{sec|Methods} we describe our methods and data analysis; in Section \ref{sec|Results} we present and discuss our results; in Section \ref{sec|Summary} we summarize our findings and outlook future perspectives. Throughout the work, we adopt the standard, flat $\Lambda$CDM cosmology with rounded parameter values \cite{Aghanim20}: matter density $\Omega_m\approx 0.3$, baryon density $\Omega_b\approx 0.05$, Hubble constant $H_0 = 100\, h$ km s$^{-1}$ Mpc$^{-1}$ with $h\approx 0.7$.

\section{Methods}\label{sec|Methods}

In this Section we recall our basics framework, with particular focus on the basic kinematic observables in dwarfs. We then discuss the data and the Bayesian analysis exploited to probe such a scenario.

\subsection{Dark Matter in Fractional Gravity}\label{sec|FracGrav}

$N$-body simulations in the standard $\Lambda$CDM cosmology indicate that virialized halos of DM particles follow an approximately universal density profile, routinely described via the Navarro-Frenk-White \cite{Navarro97} shape $\rho(r) = \rho_s\,r_s^3/r\,(r+r_s)^2$
in terms of a scale radius $r_s$ and of a characteristic density $\rho_s$.

In the standard Newtonian theory, the potential $\Phi_{\rm N}(r)$ associated to a given density distribution $\rho(r)$ can be computed from the Poisson equation:
\begin{equation}\label{eq|poisson}
\Delta\Phi_{\rm N}(\mathbf{r})=4\pi G\, \rho(\mathbf{r})
\end{equation}
where $\Delta$ is the Laplacian operator; this is an inherently local equation, in that the potential at a point depends only on the value of the density there. For the spherically symmetric NFW profile, one easily finds that
\begin{equation}\label{eq|phiN}
\Phi_{\rm N}(r) = -\frac{G M_s}{r}\,\,\ln\left(1+\frac{r}{r_s}\right)~,
\end{equation}
with $M_s\equiv 4\pi\,\rho_s\,r_s^3$. Computing the NFW mass $M(<r) = 4\pi\, \int_{0}^r{\rm d}r'\, r'^2\, \rho(r')=M_s\, [\ln (1+r/r_s)-r/(r+r_s)]$, it is easy to verify that $|{\rm d}\Phi_{\rm N}/{\rm d}r|=G\, M(<r)/r^2$, as a direct consequence of Birkhoff's theorem.

In the FG framework, the potential $\Phi_{\rm F}(r)$ is instead derived from the modified Poisson equation \cite{Giusti20a,Benetti23a}
\begin{equation}\label{eq|fracpoisson}
(-\Delta)^s\, \Phi_{\rm F} (\mathbf{r}) = -4\pi G\, \ell^{2-2s}\,\rho(\mathbf{r})
\end{equation}
where $(-\Delta)^s$ is the fractional Laplacian operator (see the excellent textbook \cite{Uchaikin13} for details), $s\in [1,3/2]$ is the fractional index (this range of values for $s$ is required to avoid divergences; see Appendix A in~\cite{Benetti23a}), and $\ell$ is a fractional length scale that must be introduced for dimensional reasons. At variance with the standard case, the fractional Laplacian is inherently nonlocal; the index $s$ measures the strength of this nonlocality, while the length scale $\ell$ can be interpreted as the typical size below which gravitational effects are somewhat reduced and above which they are instead amplified by nonlocality. 

In \cite{Benetti23a,Benetti23b} $\ell$ was left as a free parameter to be fitted by comparison with data. However, such a quantity enters only in the normalization of the potential but does not modify its radial shape; as such, it is strongly degenerate with the total mass, and very difficult to be constrained via pure kinematical data; essentially one can only infer the combination $M_s\, \ell^{2-2s}$. Therefore in the following, without loss of generality, we will set it to $\ell\approx r_s$ that is the relevant spatial scale in the NFW density. This position is equivalent to making the original Poisson equation non-dimensional in terms of quantities at $r_s$, and then fractionalize; this is a procedure often followed in the mathematical-physics literature \cite{GomezAguilar12,Ebaid19,Pavan19} to insert fractional dynamics in a system avoiding to add a dimensional parameter of ambiguous interpretation and problematic estimation. 

For $s\in [1,3/2)$, the solution reads \cite{Benetti23a}
\begin{equation}\label{eq|phiF}
\begin{aligned}
\Phi_{\rm F}(r) =&  -\frac{G M_s}{r_s}\,\frac{1}{2^{2s}\,\sqrt{\pi}}\,\frac{\Gamma\left(\frac{3}{2}-s\right)}{\Gamma(s+1)}\,\frac{r_s}{r}\,\left\{\frac{2\pi s}{\sin(2\pi s)}\,\left[\left(1+\frac{r}{r_s}\right)^{2s-2} \right.\right.\\
& \\
&-\left.\left. \left(1-\frac{r}{r_s}\right)^{2s-2}\right]+\frac{(r/r_s)^{2s}}{1-(r/r_s)^{2}}\,\left[\left(1+\frac{r}{r_s}\right)\, _{2}F_{1}\left(1,1,2s+1,\frac{r}{r_s}\right) \right.\right. \\
&\\
&+ \left.\left.\left(1-\frac{r}{r_s}\right)\, _{2} F_{1}\left(1,1,2s+1,-\frac{r}{r_s}\right)-\frac{4s}{2s-1} \right] \right\}~~~~,~~~~s\in[1,3/2)~
\end{aligned}
\end{equation}
{with $\Gamma(s) = \int_0^\infty\,{\rm d}x\, x^{s-1}\, e^{-x}$ being the Euler Gamma function and ${}_2F_1(a,b,c;x) =\sum_{k=0}^\infty$ $(a)_k\, (b)_k\, x^k/(c)_k\, k!$ being the ordinary hypergeometric function in terms of the Pochammer symbols $(q)_k$ defined as $(q)_0=1$ and $(q)_k=q\,(q+1)\,\ldots\,(q+k-1)$}; plainly, $\Phi_{\rm F}(r)$ for $s=1$ coincides with the usual expression $\Phi_{\rm N}(r)$ of Equation (\ref{eq|phiN}). For the limiting case $s=3/2$, the computation requires some principal-value regularization and the solution~reads
\begin{equation}\label{eq|PhiFlim}
\begin{aligned}
\Phi_{\rm F}(r) =& -\frac{G\,M_s}{r}\,\frac{1}{\pi}\,	
\left\{2\,\frac{r}{r_s}\, \left[\log \left(\frac{r}{r_s}\right)-1\right]-\left(1+\frac{r}{r_s}\right)\,\log \left(\frac{r}{r_s}\right)\,\log \left(1+\frac{r}{r_s}\right) \right.\\
&\\
&+\left. \left(\frac{r}{r_s}-1\right)\, \text{Li}_2\left(1-\frac{r}{r_s}\right)-\left(1+\frac{r}{r_s}\right)\, \text{Li}_2\left(-\frac{r}{r_s}\right) + \frac{\pi^{2}}{6}\right\}~~~~,~~~~~s=3/2~
\end{aligned}
\end{equation}
where ${\rm Li}_2(x)=\sum_{k=1}^\infty\,x^k/k^2$ is the dilogarithm function.

\begin{figure}[t]
\includegraphics[width=1.\textwidth]{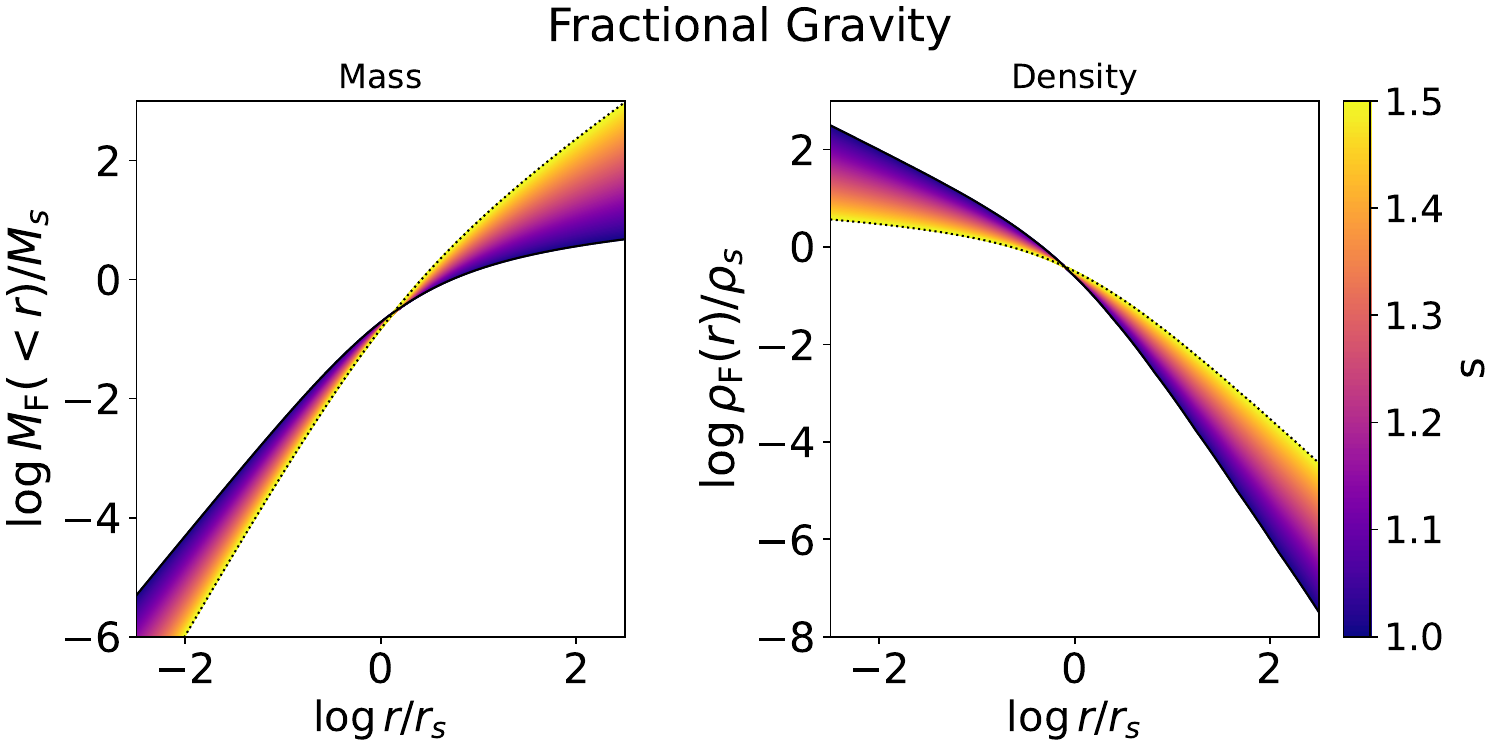}
\caption{Radial profiles of effective mass (middle) and density (right) in the FG framework for different values of the fractional index $s$ (color-coded); for reference, the dotted lines refer to the maximal value $s = 3/2$.}\label{fig|FG}
\end{figure}

Being a nonlocal framework, in FG the Birkhoff theorem does not hold, but one can insist in writing $|{\rm d}\Phi_{\rm F}/{\rm d}r| = G\, M_{F}(<r)/r^2$ in terms of an effective mass $M_{F}(<r)$; then one can differentiate the latter to obtain an effective density profile $\rho_{\rm F}(r)= (1/4\pi\,r^2) \times {\rm d} M_{\rm F}/{\rm d}r$. These are actually the mass and density profiles that one would infer by looking at the dynamical observables and interpreting them in terms of Newtonian gravity. We illustrate the effective mass and density profiles for different values of $s$ in Figure \ref{fig|FG}. With $s$ increasing from unity (Newtonian case), the mass profile steepens and the density profile flattens; in the inner region, a uniform sphere behavior (corresponding to a cored density profile) tends to be progressively enforced.


These relevant profiles depends on the NFW scale radius $r_s$ and density $\rho_s$ or equivalently the mass $M_s=4\pi\,\rho_s\,r_s^3$; however, in the following, it is convenient to trade off these quantities for the mass $M_{200}$ and the concentration $c_{200}\equiv R_{200}/r_s$ at the reference radius $R_{200}$ where the average density is $200$ times the critical density $\rho_{\rm c}$. The conversion between these variables can be performed easily from the definition of $R_{200}$ and from the NFW mass distribution. Furthermore, we adopt the relationship $c(M_{200},z)$ in the $\Lambda$CDM cosmology derived from zoom-in $N-$body simulations by \cite{Wang20} and spanning twenty orders of magnitude in DM mass within the range $M_{200}\sim 10^{-5}-10^{15}\, M_\odot$.

\subsection{Dynamical modeling}\label{sec|Dynamics}

For a rotation-dominated system, the crucial quantity to be compared with the data is the total rotation velocity, which is given by
\begin{equation}\label{eq|vrot}
v_{\rm rot}^2(r) = v_{\rm halo}^2(r)+ \cfrac{M_\star}{L}\,v_{\rm disk}^2(r)+v_{\rm gas}^2(r) 
\end{equation}
where $v_{\rm halo}^2=G\,M_{\rm F}(<r)/r$ is the contribution from the DM halo, $v_{\rm gas}^2$ is the gas contribution from HI measurements, and $v_{\rm disk}^2$ is the contribution from the disk starlight appropriately converted into the stellar mass one via a global mass-to-light ratio $M_\star/L$. The overall rotation velocity depends on three parameters, namely the stellar mass-to-light ratio $M_\star/L$, the total mass of the system $M_{200}$, and the fractional index $s$.

For a dispersion-dominated system, the crucial observable is the velocity dispersion projected along the line-of-sight (l.o.s.), which is given by \cite{Binney82,Lokas01}
\begin{equation}\label{eq|sigmalos}
\sigma_{\rm los}^2(R) = \cfrac{2}{\Sigma_\star(R)}\, \int_R^\infty{\rm d}r\,\left[1-\beta\, \cfrac{R^2}{r^2}\right]\, \cfrac{\rho_\star(r)\,\sigma_r^2(r)\, r}{\sqrt{r^2-R^2}}
\end{equation}
where  $\beta\equiv 1-\sigma_\theta^2/\sigma_r^2$ is the anisotropy parameter (hereafter assumed constant with radius) and
\begin{equation}\label{eq|densurf}
\Sigma_\star(R)=2\,\int_{R}^\infty{\rm d}r\, \cfrac{\rho_\star(r)\, r}{\sqrt{r^2-R^2}} \end{equation}
is the surface mass density of the tracers (e.g., stars) in terms of the volume one $\rho_\star(r)$; in addition, the radial velocity dispersion is obtained by solving the Jeans equation, which yields
\begin{equation}\label{eq|sigmar}
\sigma_r^2(r) = \cfrac{1}{r^{2\beta}\,\rho_\star(r)}\,\int_r^\infty{\rm d}r'\, r'^{2\beta}\, \rho_\star(r')\,\cfrac{G\, [M_{\rm F}(<r')+M_\star(<r')]}{r'^2}
\end{equation}
in terms of the tracer mass $M_\star(<r)=4\pi\int_0^r{\rm d}r'\, r'^2\,\rho_\star(r')$. 
Typically, for the dispersion-dominated galaxies considered in this work, stellar tracers are exploited, with a density distribution following the Plummer's model\footnote{For DragonFly 44 we actually exploit a Sersic surface density profile, as detailed in \cite{vanDokkum19}.} \cite{Plummer11}. In such a case $\rho_\star(r)\propto (1+r^2/r_{1/2}^2)^{-5/2}$ and $\Sigma_\star \propto (1+r^2/r_{1/2}^2)^{-2}$, with $r_{1/2}$ the 2D half-light radius; the normalization is derived from the observed surface luminosity profile via a global mass-to-light ratio $M_\star/L$. The overall dispersion velocity depends on four parameters, i.e., the mass-to-light ratio $M_\star/L$, the anisotropy parameter $\beta$, the mass of the system $M_{200}$ and the fractional index $s$. 

\subsection{Data and Bayesian Analysis}\label{sec|Bayes}

We probe the FG framework by exploiting the rotation velocity profiles of dwIrr galaxies and the l.o.s. dispersion velocity profiles of dwSph.

For rotation-dominated systems, we rely the SPARC database \cite{Lelli16,Li20}, which provides the stellar and gas contribution to the rotation velocity as found by numerically solving the standard Poisson equation for the observed surface brigthness profile at $3.6\,\mu$m (with a reference stellar mass-to-light ratio $M_\star/L=1$, so that this must be rescaled in building the total rotation velocity as in Equation \ref{eq|vrot}) and the HI surface density profile. Specifically, we consider the $8$ galaxies in SPARC that are classified as dwIrr and are flagged as having high quality data on the rotation curve: D631-7, DDO64, DDO161, UGC731, UGC5005, UGC5414, UGC7608, NGC3741. The latter is the galaxy with the best dataset, and actually constitutes the object with the most extended rotation curve (relative to the half-light radius) measured to date. 

For dispersion-dominated systems, we consider $7$ Milky-Way dwSph for which high quality determination of the spatially-resolved dispersion velocity has been obtained via stellar tracers  (\cite{Walker09,Mateo08}; see also data collection by \cite{DeMartino23} and references therein) and for which tidal effects are not appreciably influencing the inner kinematics: Carina, Leo I, Leo II, Sculptor, Draco, Sextans, and Fornax. To these we add the dispersion-dominated ultra-diffuse galaxy DragonFly 44 \cite{vanDokkum19}. Note that ultra-diffuse galaxies are a mixed bag of objects with very different properties: some feature large angular momentum, a rich gas reservoir with ongoing active star-formation activity \cite{Leisman17,ManceraPina19,ManceraPina20,Sengupta19}; others show no signs of rotation, a poor gas content and old stellar population in passive evolution \cite{vanDokkum15,vanDokkum19}. DragonFly 44 belongs to this last category, and being dispersion-dominated has been treated here along with the dwSph sample. In addition, it is a particularly interesting object since it features a very small stellar mass with respect to its large size; in fact, the DM mass is expected to dominate even at small radii. Therefore DragonFly 44 has been exploited as an useful laboratory to test the nature of DM and gravity \cite{Wasserman19,Julio23}.

\begin{table}[t]
\caption{Properties of the galaxy sample considered in this work: half-light radius, total luminosity, and mass-to-light ratio estimated from stellar population synthesis models (used in building the priors of our Bayesian analysis, see Section \ref{sec|Bayes}). Top half of the table includes rotation-dominated dwarfs, for which the listed quantities refer to the luminosity at $3.6\,\mu$m; bottom half of the Table includes dispersion-dominated dwarfs, for which the listed quantities refer to the $V-$band luminosity.}\label{tab|sample}
\newcolumntype{C}{>{\centering\arraybackslash}X}
\begin{tabularx}{\textwidth}{lCCCCCCC}
\toprule
\textbf{Galaxy} & \boldmath{$r_{1/2}$ [kpc]} & \boldmath{$\log L\, [L_    \odot]$} & \boldmath{$M_\star/L\, [M_\odot/L_\odot]$}\\
\midrule
D631-7 & $1.22$ & $8.28$ & $0.5\pm 0.1$   \\
DDO64 & $1.22$ & $8.18$ & $0.5\pm 0.1$   \\
DDO161 & $2.04$ & $8.74$ & $0.5\pm 0.1$   \\
UGC731 & $1.40$ & $8.51$ & $0.5\pm 0.1$    \\
UGC5005 & $5.0$ & $9.61$ & $0.5\pm 0.1$   \\
UGC5414 & $2.33$ & $9.05$ & $0.5\pm 0.1$   \\
UGC7608 & $1.60$ & $8.42$ & $0.5\pm 0.1$  \\
NGC3741 & $0.32$ & $7.45$ & $0.5\pm 0.1$   \\
\midrule
Carina & $0.27$ & $5.57$ & $3.4\pm 2.9$  \\
Leo I & $0.29$ & $6.74$ & $8.8\pm 5.6$  \\
Leo II & $0.22$ & $5.87$ & $0.4\pm 0.4$  \\
Sculptor & $0.31$ & $6.36$ & $3.6\pm 2.0$  \\
Draco & $0.24$ & $5.45$ & $11.1\pm 4.7$  \\
Sextans & $0.75$ & $5.64$ & $8.5\pm 3.3$  \\
Fornax & $0.79$ & $7.31$ & $7.1\pm 6.0$  \\
DragonFly 44 & $3.87$ & $8.37$ & $1.5\pm 0.4$  \\
\bottomrule
\end{tabularx}
\end{table}

In Table \ref{tab|sample} we report some relevant properties of the galaxies considered in our analysis. Specifically, the first two columns list the circularized half-light radius of the projected surface brightness profile and the total luminosity as determined from photometric observations (uncertainties are negligible for the purpose of this analysis); these quantities refer to the $3.6\,\mu$m band for dwIrr and to the $V-$band for dwSph. The third column lists the stellar mass-to-light-ratio expected from stellar population synthesis models \cite{Bell01,Portinari04,Schombert19}: for disk-dominated dwIrr values $M_\star/L\approx 0.5$ applies with little uncertainties at $3.6\, \mu$m; for dwSph the $M_\star/L$ values are estimated from the $V-I$ color index and thus are more dispersed and uncertain \cite{DeMartino23,Wasserman19}.

For our Bayesian analysis, we consider the parameter set $\theta\equiv (M_\star/L,M_{200},s)$ for rotation-dominated dwarfs and $\theta\equiv (M_\star/L,\beta,M_{200},s)$ for dispersion-dominated ones. To estimate these parameters, we adopted a Bayesian framework and built the Gaussian log-likelihood
\begin{equation}\label{eq|likelihood}
\log \mathcal{L}(\theta) = -\chi^2(\theta)/2~.
\end{equation}
where the chi-square $\chi^2(\theta)=\sum_{i} [\mathcal{M}(\theta,r_i)-\mathcal{D}(r_i)]^2/\sigma_{\mathcal{D}}^2(r_i)$ is obtained by comparing our empirical model expectations $\mathcal{M}(\theta,r_i)$ to the data values $\mathcal{D}(r_i)$ with their uncertainties $\sigma_{\mathcal{D}}(r_i)$, summing over the different radial coordinates $r_i$ of the data.

We adopt flat priors $\pi(\theta)$ on $s\in [1,3/2]$ and on $\log M_{200}\, [M_\odot]\in [6,13]$. Moreover, we assume a lognormal prior on $\log M_\star/L$ with average and dispersion as expected from stellar population synthesis models (see Table \ref{tab|sample}).
As to $\beta$, since by definition it varies in the range $(-\infty,1]$, we actually prefer to perform inference on the symmetrized version $\beta_{\rm sym}\equiv \beta/(2-\beta)$ that maps the original quantity in a compact domain $\beta_{\rm sym}\in (-1,1]$; a flat prior on $\beta_{\rm sym}$ within this range is used. Finally, to help robustly break any possible degeneracy between the halo and stellar masses, we follow \cite{Li20} and add as a $\Lambda$CDM prior the stellar mass vs. halo mass relation derived from multi-epoch abundance matching by \cite{Moster13}, which is also consistent with independent observational determinations from satellite kinematics \cite{Lange19}, rotation curve modeling \cite{Lapi18}, and weak lensing analysis \cite{Mandelbaum16}.

We then sample the parameter posterior distributions $\mathcal{P}(\theta) \propto \mathcal{L}(\theta)\,\pi(\theta)$ via the MCMC Python package \texttt{emcee} \cite{ForemanMackey13}, running it with $10^4$ steps and $100$ walkers; each walker is initialized with a random position extracted from the priors discussed above. To speed up convergence, we adopt a mixture of differential evolution and snooker moves of the walkers, in proportion of $0.8$ and $0.2$ respectively, that emulates a parallel tempering algorithm. After checking the auto-correlation time, we remove the first $30\%$ of the flattened chain to ensure burn-in; the typical acceptance fractions of the various runs are around $30\%$.

\begin{table}[t]
\caption{Marginalized posterior estimates (mean and $1\sigma$ confidence intervals are reported) for the parameters from the MCMC analysis of the individual rotation-dominated dwIrr in fractional and Newtonian gravity (marked by $s=1$). Columns report the values of the stellar mass-to-light ratio $M_\star/L$, of the DM mass $M_{200}$, of the fractional index $s$, of the reduced $\chi_r^2$ for the overall fit, and of the Bayesian inference criterion (BIC) for model comparison.}\label{tab|results_dwIrr}
\newcolumntype{C}{>{\centering\arraybackslash}X}
\begin{tabularx}{\textwidth}{lCCCCCCCCCC}
\toprule
\textbf{Galaxy} & \boldmath{$\log M_\star/L$} \boldmath{[$M_\odot/L_\odot$]} &\boldmath{$\log M_{200}$} \boldmath{$[M_\odot]$} & \boldmath{$s$} & \boldmath{$\chi_r^2$} & \boldmath{BIC}\\
\midrule
&&&&\\
D631-7 & $-0.39^{+0.07}_{-0.07}$ & $10.59^{+0.06}_{-0.06}$ & $>1.47$ & $4.25$ & $63$ \\
& $-0.77^{+0.07}_{-0.07}$ & $10.15^{+0.03}_{-0.03}$ &$1$ & $19.50$ & $279$\\
&&&&\\
DDO64 & $-0.28^{+0.09}_{-0.09}$ &$10.55^{+0.07}_{-0.07}$  &$1.16^{+0.04}_{-0.04}$ & $0.56$ & $14$ \\
& $-0.45^{+0.08}_{-0.08}$ & $10.36^{+0.06}_{-0.06}$ &$1$ & $1.66$ & $25$\\
&&&&\\
DDO161 & $-0.76^{+0.07}_{-0.07}$ & $10.26^{+0.03}_{-0.02}$ &$1.35^{+0.02}_{-0.02}$ & $0.75$ & $31$  \\
& $-0.82^{+0.06}_{-0.06}$ & $10.44^{+0.01}_{-0.01}$ &$1$ & $15.27$ & $450$\\
&&&&\\
UGC731 & $-0.26^{+0.08}_{-0.08}$ & $10.70^{+0.03}_{-0.03}$ &$1.05^{+0.02}_{-0.03}$ &$0.30$ & $12$   \\
& $-0.29^{+0.08}_{-0.08}$ & $10.67^{+0.02}_{-0.02}$ &$1$ & $0.66$ & $14$ \\
&&&&\\
UGC5005 & $-0.44^{+0.08}_{-0.08}$ & $11.00^{+0.06}_{-0.06}$  &$1.29^{+0.06}_{-0.06}$ & $0.64$ & $15$ \\
& $-0.47^{+0.08}_{-0.08}$ & $10.98^{+0.05}_{-0.05}$ &$1$ & $3.13$ & $35$\\
&&&&\\
UGC5414 & $-0.28^{+0.09}_{-0.09}$ & $10.91^{+0.07}_{-0.07}$  &$1.28^{+0.04}_{-0.06}$ & $0.42$ & $8$ \\
& $-0.67^{+0.07}_{-0.07}$ & $10.54^{+0.04}_{-0.04}$ &$1$ & $7.37$ & $34$ \\
&&&&\\
UGC7608 & $-0.25^{+0.09}_{-0.09}$ & $10.68^{+0.07}_{-0.07}$ &$1.08^{+0.03}_{-0.06}$ & $1.26$ & $14$ \\
& $-0.29^{+0.09}_{-0.09}$ & $10.63^{+0.06}_{-0.06}$ &$1$ & $1.35$ & $14$\\
&&&&\\
NGC3741 & $-0.21^{+0.08}_{-0.08}$ & $10.29^{+0.04}_{-0.04}$ &$1.23^{+0.03}_{-0.03}$ & $0.47$ & $20$\\
& $-0.49^{+0.07}_{-0.07}$ & $10.09^{+0.02}_{-0.02}$ &$1$ & $4.65$ & $97$\\
&&&&\\
\bottomrule
\end{tabularx}
\end{table}

\section{Results}\label{sec|Results}

\begin{center}
\begin{figure}[H]
\includegraphics[width=0.45\textwidth]{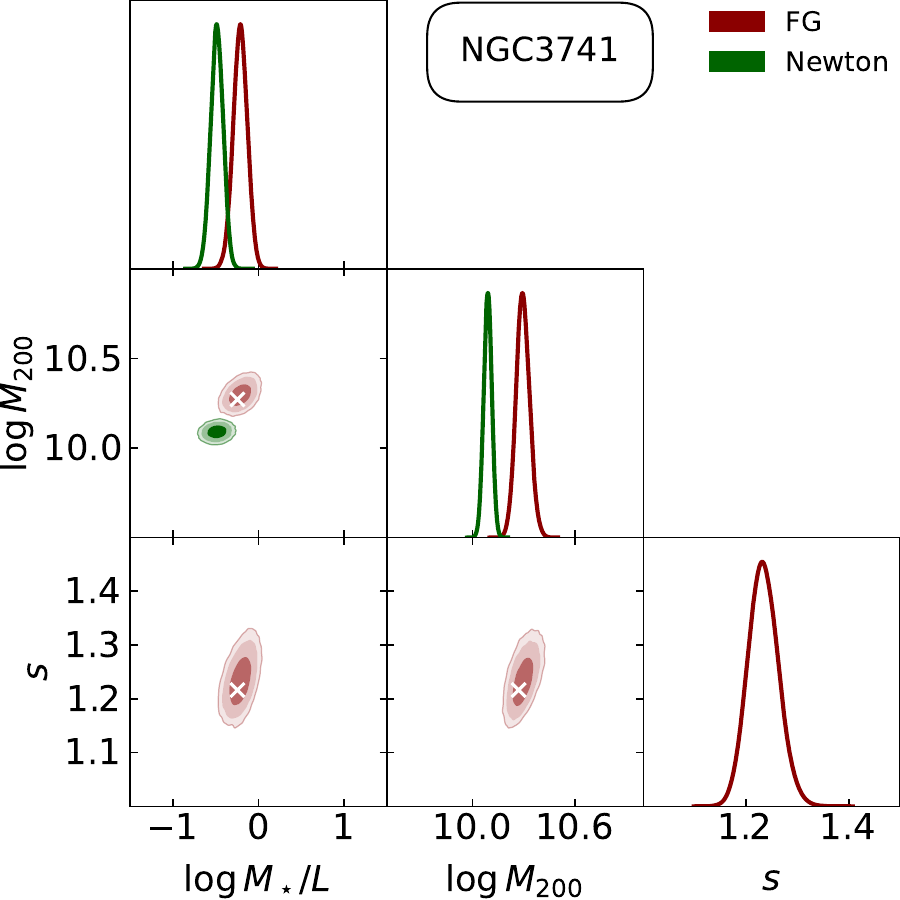}~~~~~~~~\includegraphics[width=0.45\textwidth]{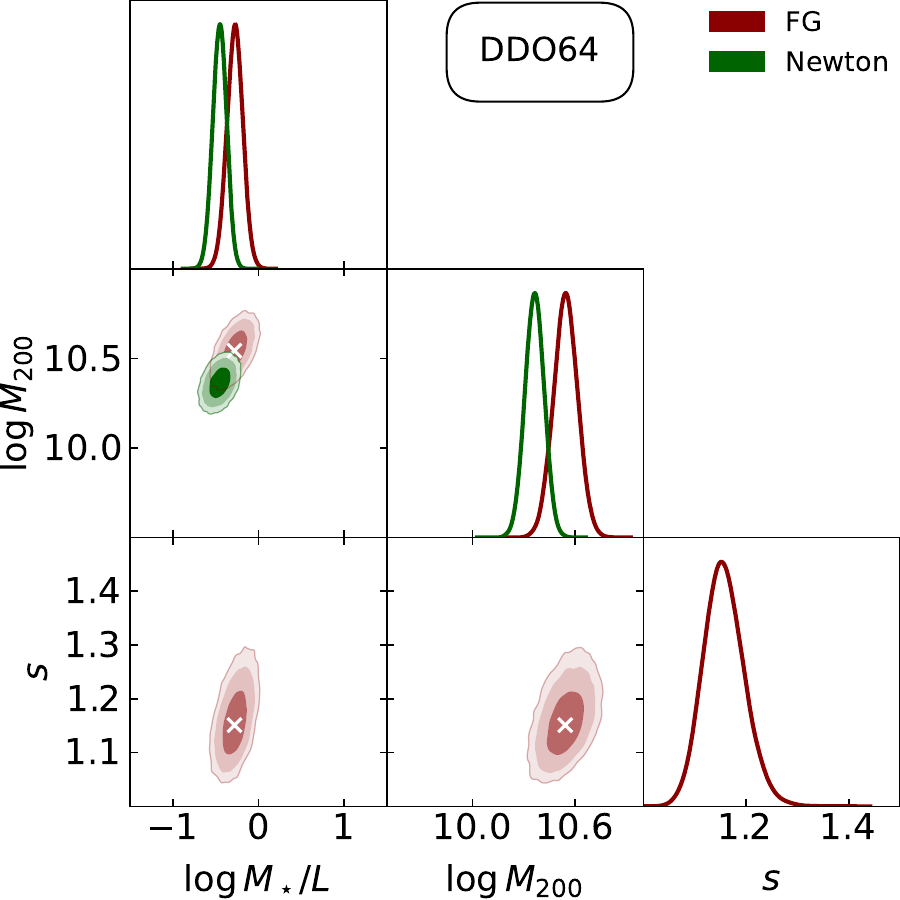}\\
\includegraphics[width=0.5\textwidth]{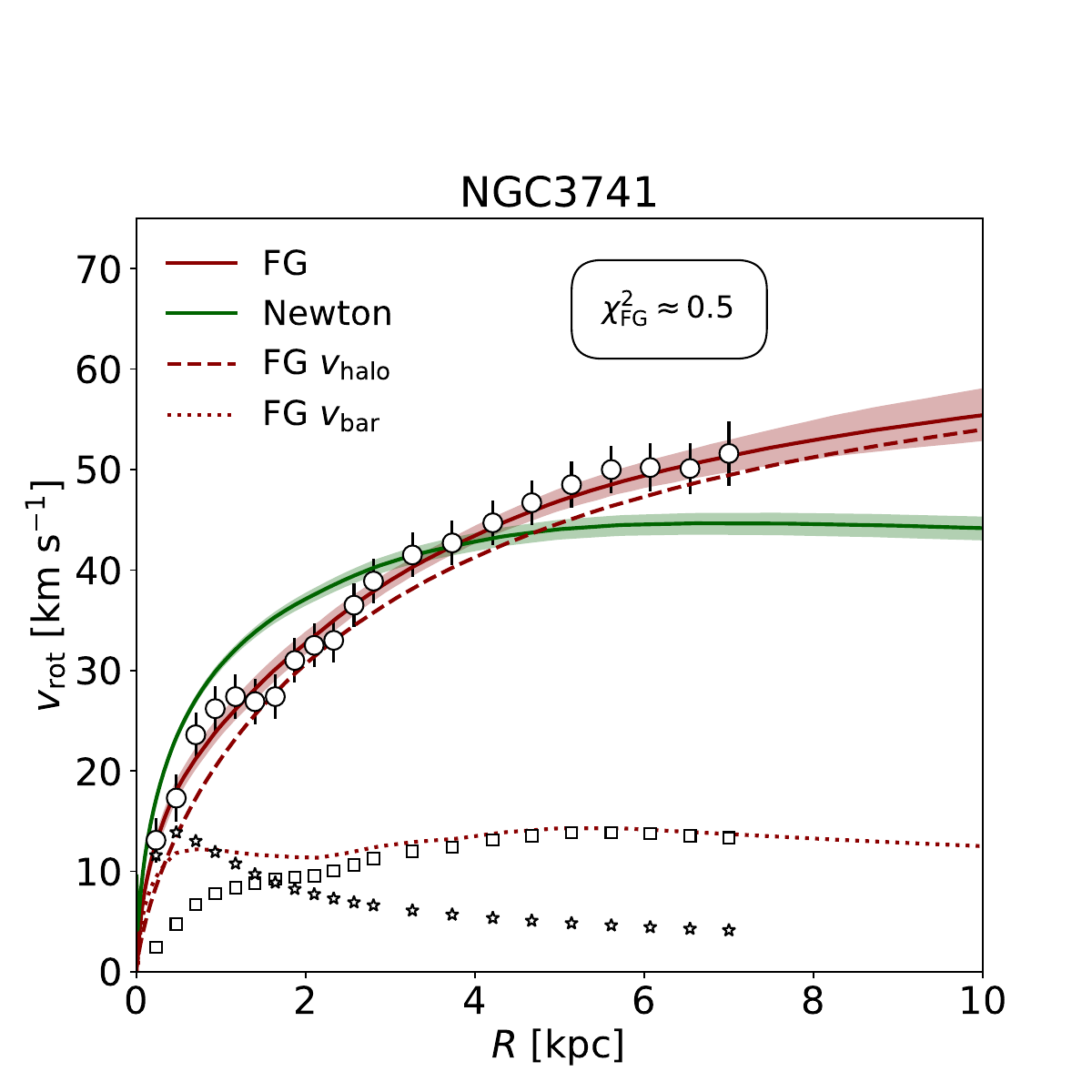}
\includegraphics[width=0.5\textwidth]{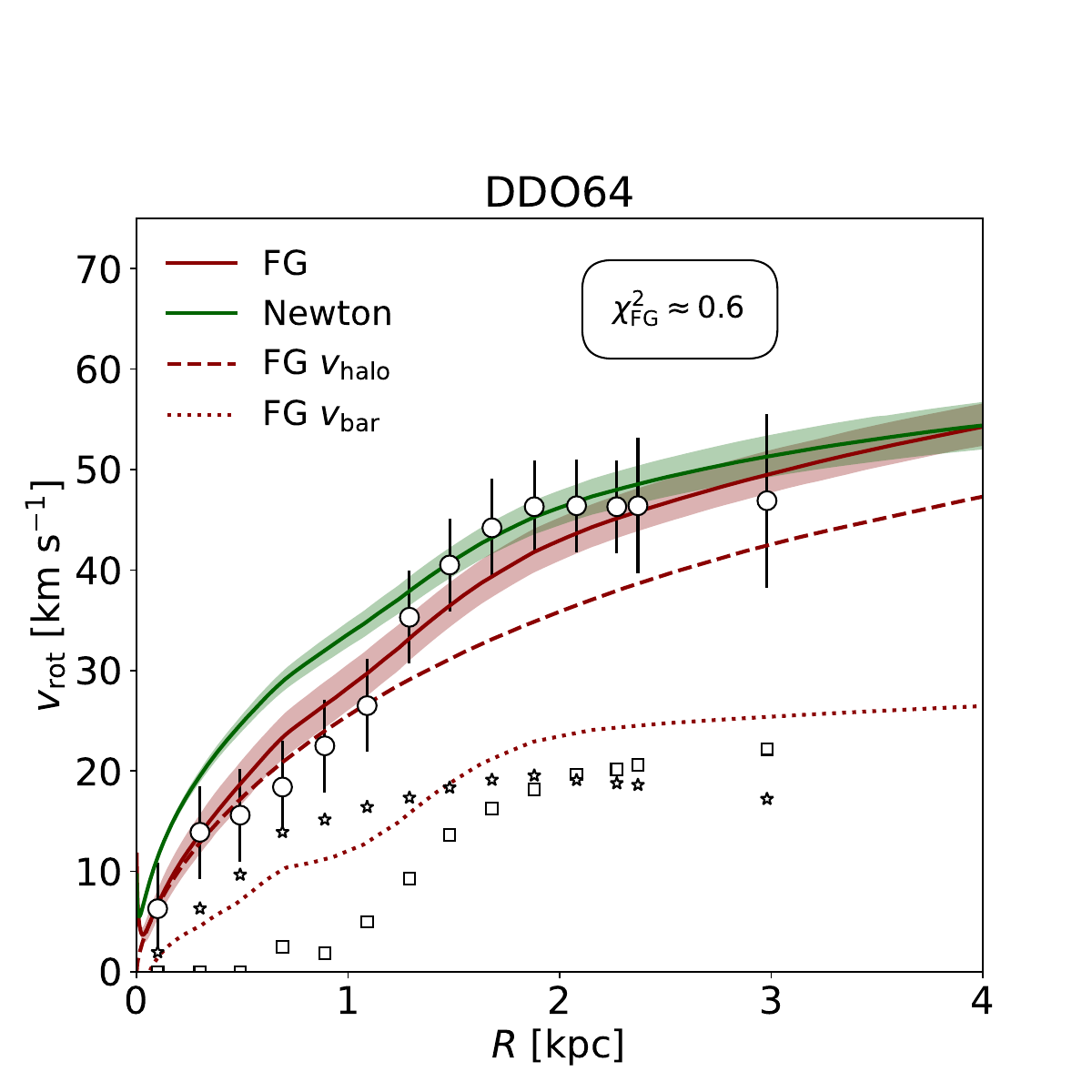}
\caption{Top panel: MCMC posterior distributions of the stellar mass-to-light ratio $M_\star/L$, the DM mass $M_{200}$ and the fractional index $s$ for the galaxy NGC3741 (left) and DDO64 (right). Colored contours/lines refer to the standard Newtonian (green) and to the FG framework (red). The contours show $1-2-3\sigma$ confidence intervals, with the bestfit values in FG identified by white crosses. The marginalized distributions are in arbitrary units (normalized to 1 at their maximum value). Bottom panel: Fits to the rotation curve with the Newtonian (green) and the FG (red) framework for the galaxy NGC3741 (left) and DDO64 (right). Solid lines refer to the total rotation velocity, while (for clarity only in FG) the dashed line highlights the halo contribution and the dotted line the baryonic one.  Solid lines illustrate the median, and the shaded areas show the $2\sigma$ credible interval from sampling the posterior distribution. The value of the reduced $\chi^2_r$ of the fit for FG is also reported. Circles represent data from the SPARC database \cite{Lelli16} for the total rotation curve, while the contributions from the stellar (for $M_\star/L=1$) and gaseous disk are highlighted by the starred and squared symbols, respectively.}\label{fig|NGC3741}
\end{figure}
\end{center}

\begin{figure}[H]
\includegraphics[width=1.\textwidth]{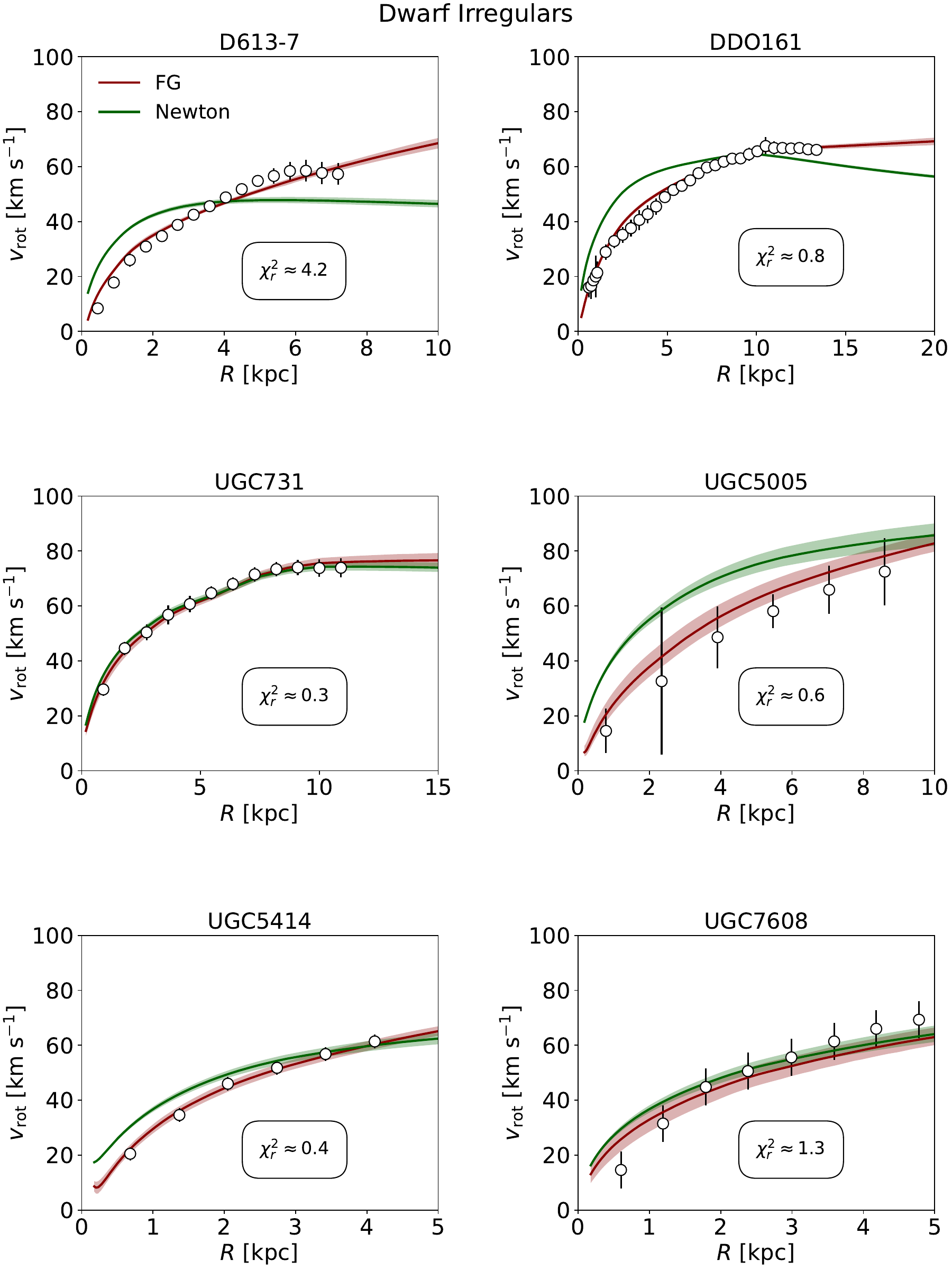}
\caption{Same as bottom panel in the previous figures for other $6$ dwIrr galaxies, as labeled. For clarity, data and models only for the total rotation curves are shown.}\label{fig|dwIrr}
\end{figure}

The results of our Bayesian analysis for rotation-dominated dwIrr galaxies are displayed in Figures \ref{fig|NGC3741}, \ref{fig|dwIrr} and in Table \ref{tab|results_dwIrr}. Specifically, in the top panel of Figure \ref{fig|NGC3741} we illustrate the MCMC posterior distributions for two representative dwIrr in the sample, namely NGC3741 (the one with the best and most extended data) and DDO64. Red lines/contours refer to the outcomes for FG, and green ones for Newtonian gravity; the white crosses mark the best-fit value of the parameters in FG. In the bottom panel the best fit (solid lines) and the $2\sigma$ credible intervals (shaded areas) sampled from the posteriors are shown. The solid line is for the total rotation velocity, while the dashed and dotted lines show the halo and disk contribution, respectively; for comparison, the Newtonian bestfit to the total velocity is reported in green. In Figure \ref{fig|dwIrr} the bestfits in FG and in the Newtonian case are illustrated for the other $6$ dwIrr in the sample. In Table \ref{tab|results_dwIrr} we summarize the marginalized posterior estimates of the parameters, both in FG and in the standard Newtonian case (marked with $s=1$). Columns report the median values and the $1\sigma$ credible intervals of the stellar mass-to-light ratio $M_\star/L$, of the DM mass $M_{200}$, and of the fractional index $s$; the reduced $\chi_r^2$ of the fit, and the Bayesian inference criterion (BIC) for model comparison\footnote{The BIC is defined as BIC $= 2\,\ln{\mathcal{L}_{\rm max}}+ N_{\rm par}\,\ln{N_{\rm data}}$ in terms of the maximum likelihood estimate $\mathcal{L}_{\rm max}$, of the number of parameters $N_{\rm par}$, and the number of data points $N_{\rm data}$. The BIC comes from approximating the Bayes factor, which gives the posterior odds of one model against another, presuming that the models are equally favored a priori. Note that what matters is the relative value of the BIC among different models; in particular, a difference of around ten or more indicates evidence in favor of the model with the smaller value.} are also reported.

The FG fits are always excellent, and comparable or better then those in Newtonian gravity. In particular, for D631-7, DDO64, DDO161, UGC5005, UGC5414, and NGC3741 there is a clear preference for FG both in terms of $\chi_r^2$ and of the BIC. In such cases, the fractional index takes on typical values $s\approx 1.2-1.3$, the stellar mass-to-light ratios $M_\star/L$ are slightly larger than for the Newtonian case and more in line with the value around $0.5$ expected from stellar population synthesis models, and the DM masses $M_{200}$ are appreciably larger of factors a few than in the Newtonian fit. In the other cases, namely UGC731 and UGC7608, $s$ is close to $1$, the estimates of the fitting parameters in FG and in the Newtonian setting are consistent to within $1\sigma$, and the overall quality of the fits are comparable. We have looked for some property of these two galaxies that could correlate with their smaller values of $s$, but were unable to reach a definite conclusion. Maybe an interesting evidence comes from the kinematics of their HI disks, which appear asymmetric and disturbed by a past or ongoing gravitational interaction \cite{Swaters02}; this could possibly alters the shape of the outer rotation curve and originate a variant outcome when modeling it in the FG framework. An extended sample of dwIrr with high quality
rotation curve and environmental characterization would be needed to investigate further the issue in a statistically sound manner.

As it can be seen in the top panel of Figure \ref{fig|NGC3741} for the representative cases of NGC3741 and DDO64, in FG there is no strong degeneracy in the fitting parameters, besides a weak direct dependence between $s$ and both $M_\star/L$ and $M_{200}$. The bottom panel of the same Figure shows that for NGC3741 and DDO64 the halo component largely dominates the rotation curve, with the baryonic contribution being relevant only in the innermost region within a few $r_{1/2}$; this situation is shared by all dwIrr in the analyzed sample. Therefore the shape of the rotation curve strongly constrains the halo mass/density profile; in particular, the rising trend of the rotation velocity out to large radii is difficult to be reproduced with a NFW profile in standard gravity, while the task can be easily achieved in the FG framework. 

\begin{table}[t]
\caption{Marginalized posterior estimates (mean and $1\sigma$ confidence intervals are reported) for the parameters from the MCMC analysis of the individual dispersion-dominated dwarf spheroidal in fractional (first lines) and Newtonian (second lines, with $s=1$) gravity. Columns report the values of the stellar mass-to-light ratio $M_\star/L$, of the symmetrized anisotropy parameter $\beta_{\rm sym}$, of the DM mass $M_{200}$, of the fractional index $s$, of the reduced $\chi_r^2$ for the overall fit, and of the Bayesian inference criterion (BIC) for model comparison.}\label{tab|results_dwSph}
\newcolumntype{C}{>{\centering\arraybackslash}X}
\begin{tabularx}{\textwidth}{lCCCCCCCCC}
\toprule
\textbf{Galaxy} & \boldmath{$\log M_\star/L$} \boldmath{$[M_\odot/L_\odot]$} &   \boldmath{$\beta_{\rm sym}$} & \boldmath{$\log M_{200}$} \boldmath{$[M_\odot]$} & \boldmath{$s$} & \boldmath{$\chi_r^2$} & \boldmath{BIC}\\
\midrule
&&&&\\
Carina & $+0.54^{+0.26}_{-0.26}$ & $+0.19^{+0.11}_{-0.11}$ & $9.77^{+0.12}_{-0.12}$ & $1.37^{+0.04}_{-0.05}$ & $1.67$ & $48$\\
 & $-0.75^{+0.28}_{-0.16}$ & $-0.81^{+0.013}_{-0.12}$ & $9.06^{+0.13}_{-0.06}$ & $1$ & $3.49$ & $87$\\
\\ 
Leo I & $+0.74^{+0.12}_{-0.07}$ & $-0.08^{+0.16}_{-0.16}$ & $10.35^{+0.08}_{-0.07}$ & $1.37^{+0.11}_{-0.06}$ & $0.64$& $17$\\
 & $+0.19^{+0.16}_{-0.12}$ & $-0.80^{+0.05}_{-0.07}$ & $10.03^{+0.09}_{-0.07}$ & $1$ & $2.64$ & $39$\\
\\
Leo II & $-0.40^{+0.28}_{-0.28}$ & $+0.37^{+0.25}_{-0.25}$ & $9.50^{+0.13}_{-0.13}$ & $1.11^{+0.04}_{-0.05}$ & $0.37$& $11$\\
 & $-0.74^{+0.24}_{-0.24}$ & $+0.09^{+0.27}_{-0.27}$ & $9.32^{+0.11}_{-0.11}$ & $1$ & $0.42$ & $9$\\
\\
Sculptor & $+0.60^{+0.19}_{-0.19}$ & $-0.04^{+0.06}_{-0.06}$ & $10.13^{+0.09}_{-0.09}$ & $1.23^{+0.03}_{-0.05}$ & $1.32$ & $54$\\
 & $-0.37^{+0.13}_{-0.13}$ & $-0.32^{+0.06}_{-0.06}$ & $9.51^{+0.06}_{-0.06}$ & $1$ & $1.65$ & $65$\\
\\
Draco & $+1.05^{+0.14}_{-0.14}$ & $+0.01^{+0.17}_{-0.17}$ & $9.93^{+0.09}_{-0.09}$ & $1.13^{+0.03}_{-0.03}$ & $0.27$ & $14$\\
 & $+0.75^{+0.12}_{-0.12}$ & $-0.32^{+0.24}_{-0.17}$ & $9.71^{+0.07}_{-0.07}$ & $1$ & $0.56$ & $15$\\
\\
Sextans & $+0.90^{+0.13}_{-0.13}$ & $-0.34^{+0.07}_{-0.05}$ & $9.94^{+0.08}_{-0.08}$ & $>1.45$ & $1.54$ & $24$\\
 & $+0.54^{+0.12}_{-0.12}$ & $-0.89^{+0.03}_{-0.07}$ & $9.62^{+0.07}_{-0.07}$ & $1$ & $3.07$ & $38$\\
\\
Fornax & $-0.36^{+0.15}_{-0.12}$ & $+0.16^{+0.03}_{-0.03}$ & $10.03^{+0.1}_{-0.08}$ & $>1.49$ & $1.30$ & $69$\\
 & $<-1.84$ & $-0.16^{+0.04}_{-0.04}$ & $9.08^{+0.03}_{-0.03}$ & $1$ & $1.15$ & $62$\\
\\
DragonFly 44 & $+0.17^{+0.09}_{-0.09}$ & $-0.12^{+0.11}_{-0.07}$ & $10.79^{+0.07}_{-0.07}$ & $>1.31$ & $0.46$& $12$\\
 & $+0.09^{+0.09}_{-0.09}$ & $-0.48^{+0.08}_{-0.08}$ & $10.69^{+0.06}_{-0.06}$ & $1$ & $1.27$ & $16$\\
&&&&\\
\bottomrule
\end{tabularx}
\end{table}

\begin{center}
\begin{figure}[H]
\includegraphics[width=0.475\textwidth]{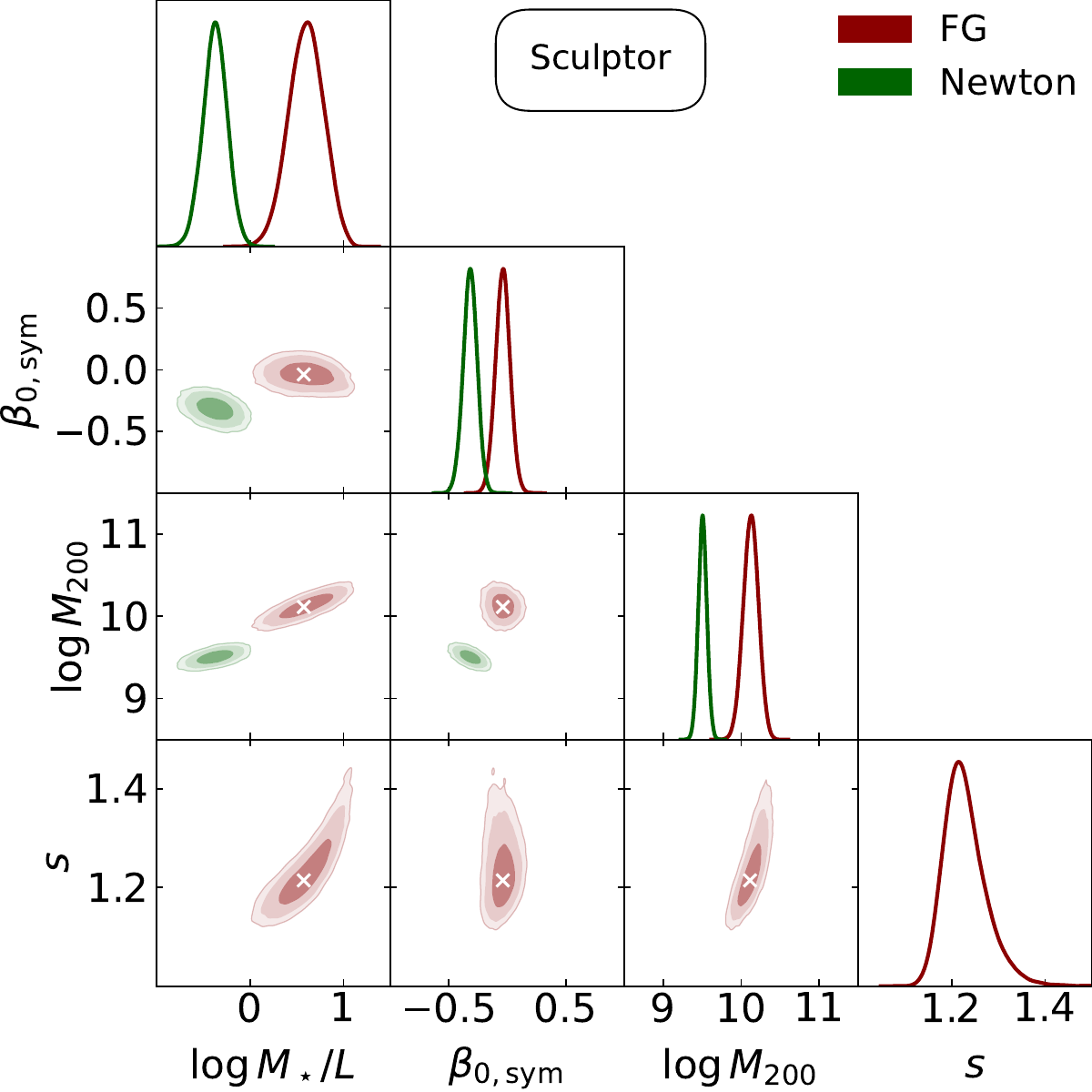}~~~~~~~~\includegraphics[width=0.475\textwidth]{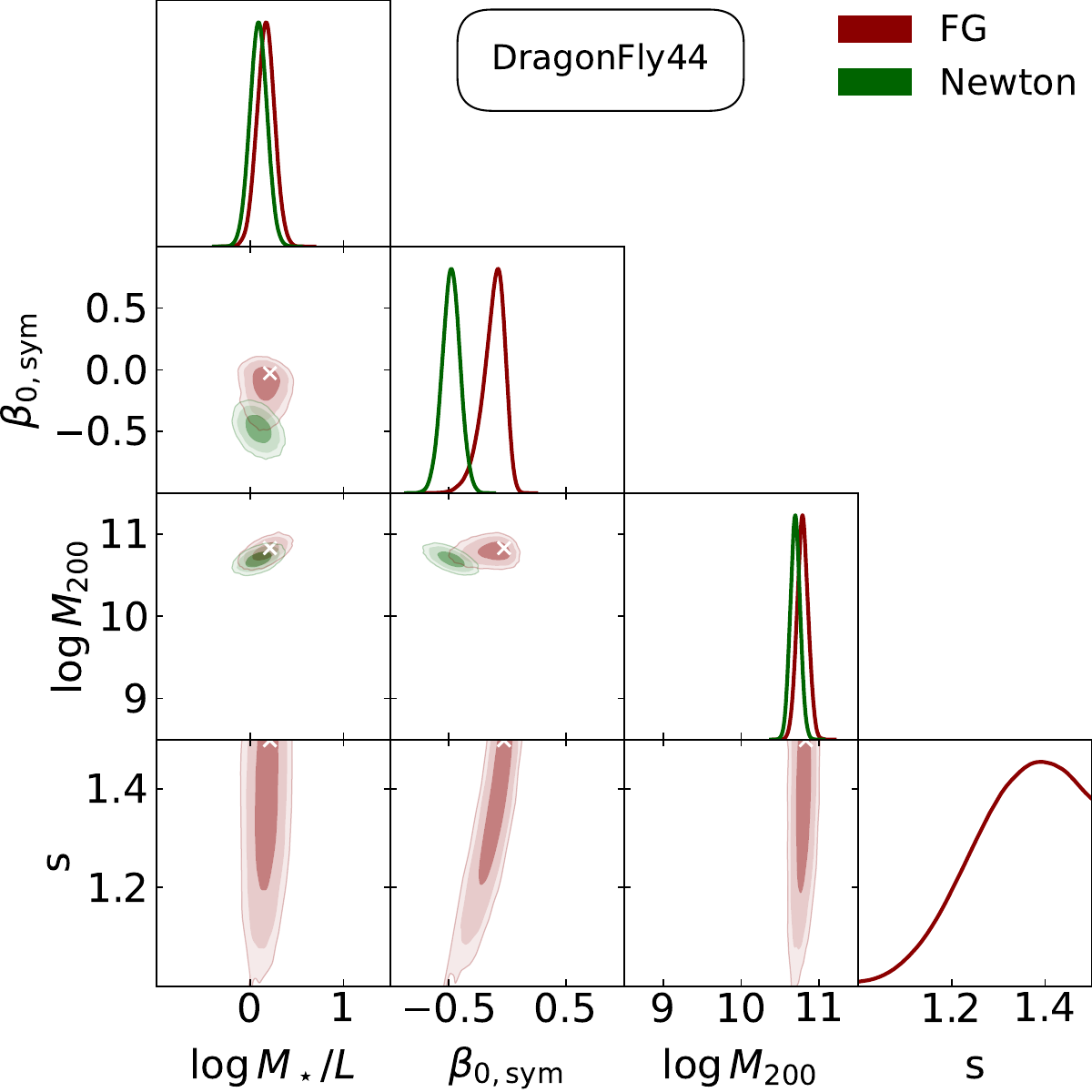}\\
\includegraphics[width=0.525\textwidth]{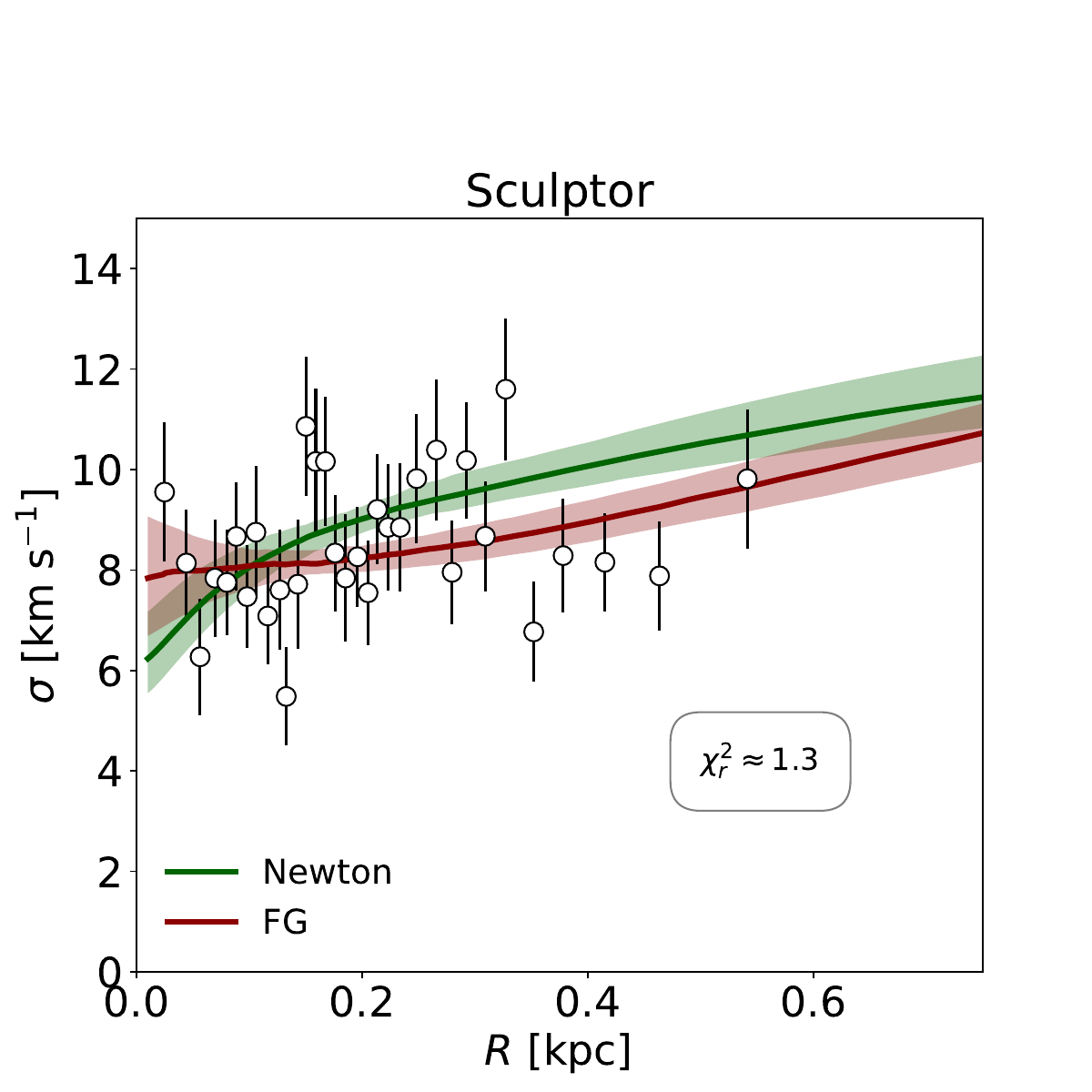}
\includegraphics[width=0.525\textwidth]{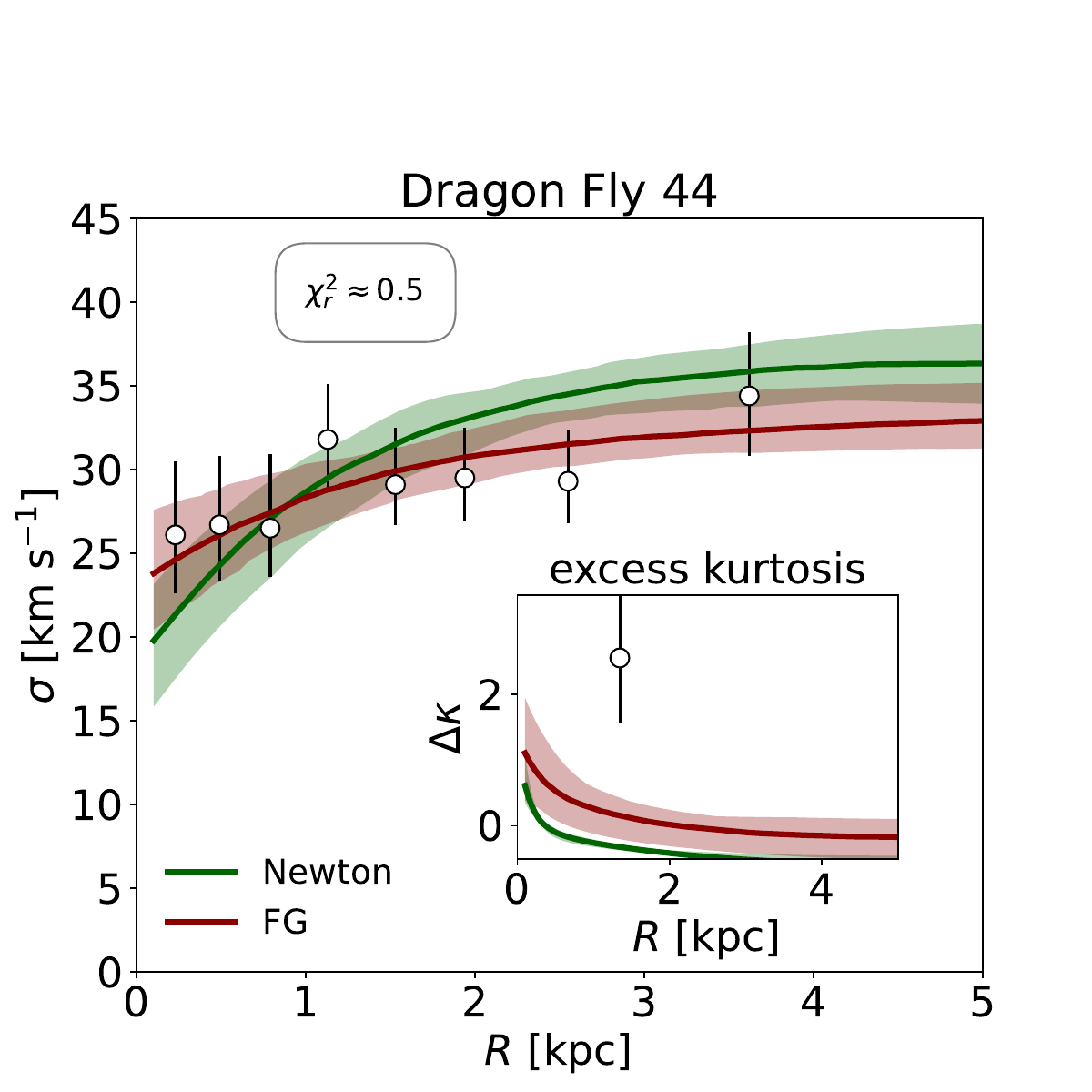}
\caption{Top panel: MCMC posterior distributions of the mass-to-light ratio $M_\star/L$, the symmetrized anisotropy parameter $\beta_{\rm sym}$, the mass $M_{200}$ and the fractional index $s$ for the dwSph galaxy Sculptor (left) and the ultra-diffuse galaxy DragonFly 44 (right). Colored contours/lines refer to the standard Newtonian (green) and to the FG framework (red). The contours show $1-2-3\sigma$ confidence intervals, with the bestfit values in FG identified by white crosses. The marginalized distributions are in arbitrary units (normalized to 1 at their maximum value). Bottom panel: Fits to the l.o.s. dispersion profile with the Newtonian (green) and the FG (red) framework for the dwSph galaxy Sculptor (left) and DragonFly 44 (right). The inset on the right bottom panel illustrates the excess kurtosis $\Delta\kappa$ with respect to a Gaussian velocity distribution. Solid lines illustrate the median, and the shaded areas show the $2\sigma$ credible interval from sampling the posterior distribution. The value of the reduced $\chi^2_r$ of the fit for FG is also reported. Circles represent data from \cite{Walker09} for Sculptor and from \cite{vanDokkum19} for DragonFly 44.}\label{fig|Dragonfly44}
\end{figure}
\end{center}

\begin{figure}[H]
\includegraphics[width=1.\textwidth]{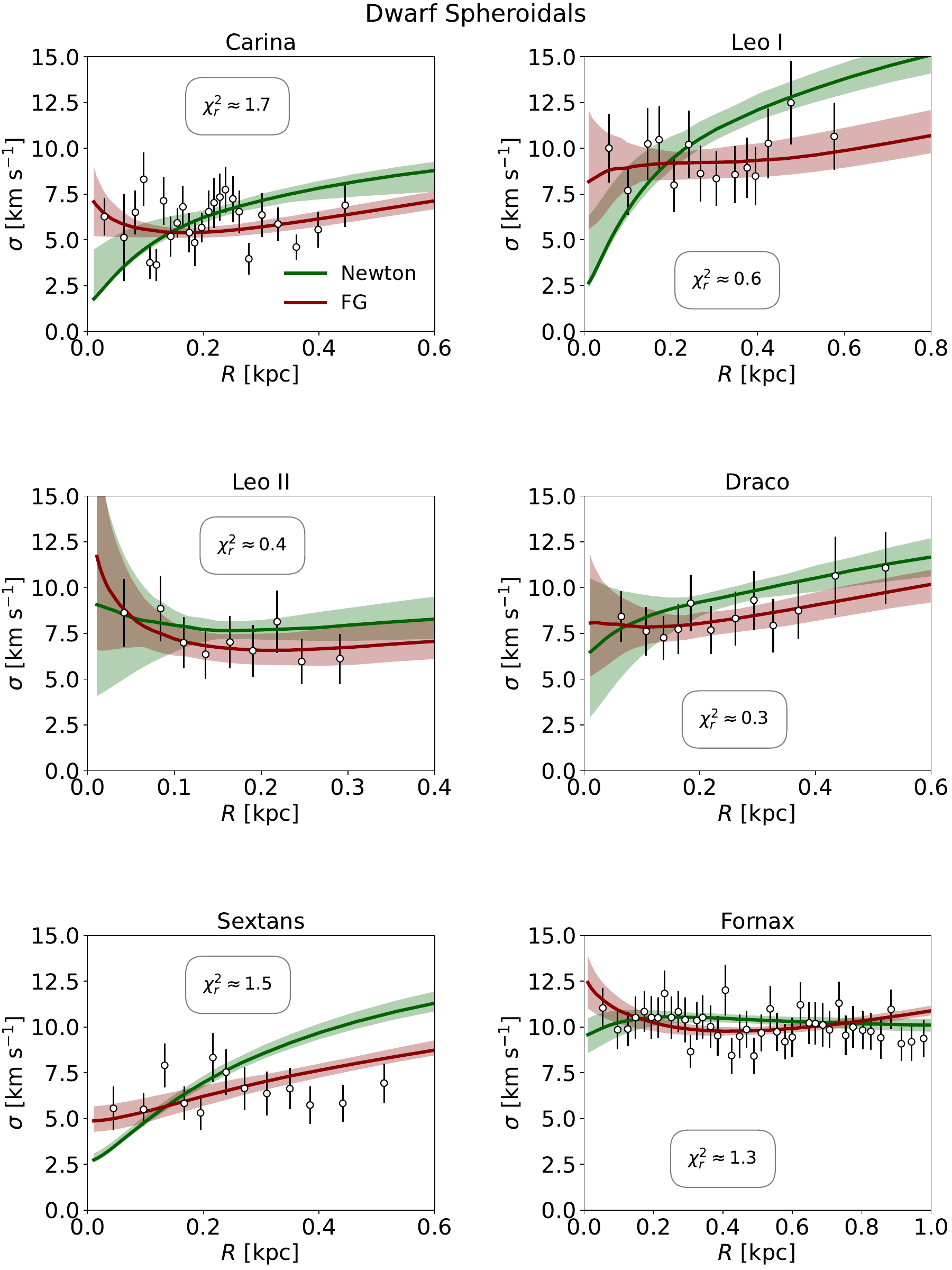}
\caption{Same as bottom panel in the previous figure for $6$ dwSph galaxies, as labeled.}\label{fig|dwSph}
\end{figure}

The results for dispersion-dominated galaxies are displayed in Figures \ref{fig|Dragonfly44}, \ref{fig|dwSph} and in Table \ref{tab|results_dwSph}. Figure \ref{fig|Dragonfly44} focuses on the representative dwSph Sculptor and on the ultra-diffuse galaxy DragonFly 44. In the top panel of Figure \ref{fig|Dragonfly44} we illustrate the MCMC posterior distributions for Sculptor and DragonFly 44. As above, red lines/contours refer to the outcomes for FG, and green ones for Newtonian gravity, with the white crosses marking the best-fit value of the parameters in FG. In the bottom panel the best fit (solid lines) and the $2\sigma$ credible intervals (shaded areas) sampled from the posteriors are shown, with the reference Newtonian fit in green. In Figure \ref{fig|dwSph} the fits in FG and in the Newtonian case are illustrated for the other $6$ dwSph in the sample. In Table \ref{tab|results_dwSph} we summarize the marginalized posterior estimates of the parameters for dwSph, both in FG and in the standard Newtonian case (marked with $s=1$). Columns report the median values and the $1\sigma$ credible intervals of the stellar mass-to-light ratio $M_\star/L$, of the symmetrized anisotropy parameter $\beta_{\rm sym}$, of the DM mass $M_{200}$, and of the fractional index $s$; the reduced $\chi_r^2$ of the fit, and the BIC are also reported.

The FG fits to dispersion-dominated galaxies are very good, and in several instances appreciably better than in Newtonian gravity. In particular, for Carina, Leo I, Sculptor, and Sextans there is a clear preference for FG both in terms of $\chi_r^2$ and in terms of the BIC. In such cases, the fractional index takes on typical values $s\gtrsim 1.2$, the stellar mass-to-light ratios $M_\star/L$ are appreciably larger than for the Newtonian case and more in line with the prior from stellar population synthesis models, and the DM masses $M_{200}$ are substantially larger of factors several with respect to the Newtonian fits. In other cases, namely Leo II and Draco, the index $s\lesssim 1.1$ is smaller, the estimates of the fitting parameters in FG and in the Newtonian setting are consistent to with $3\sigma$, and the overall quality of the fits are comparable. Finally, in the cases of Fornax and DragonFly 44 there is a clear preference for large values of $s\approx 1.5$, but the improvement in the FG fits with respect to the Newtonian case, albeit clear at a visual inspection, is not statistically significant enough to make definite conclusions. 

As it can be seen in the top panel of Figure \ref{fig|Dragonfly44} for Sculptor and DragonFly 44, the most relevant degeneracy between the fitting parameters involves $s$ and the $\beta_{\rm sym}$, in such a way that FG models with larger $s$ tend to be more isotropic. In fact, for Sculptor and DragonFly 44 the FG fit shows preference for almost isotropic orbits, while the Newtonian fit favors tangentially-dominated motions. The bottom panel of the same Figure illustrates visually the quality of the FG fit for Sculptor and DragonFly 44, which is excellent within the scatter of the datapoints. FG performs definitely better than the Newtonian case for Sculptor, while for DragonFly 44 the evidence is made barely significant in terms of reduced $\chi_r^2$ and of the BIC. In this respect, however, it is also interesting to look at the inset where the excess kurtosis $\Delta \kappa$ (the kurtosis is related to the fourth velocity moments of the stellar tracers, and the excess is respect to the value $3$ for a reference Gaussian velocity distribution) is illustrated. Although the measured value is largely uncertain, there is clear a tendency for a definite positive $\Delta\kappa$; qualitatively, this is consistent with the FG result at $2\sigma$, while being highly discordant (more than $3\sigma$) with the Newtonian fit. It is worth mentioning that the estimated $M_{200}\lesssim 10^{11}\, M_\odot$ from our analysis in FG is consistent with the recent determination from the abundance of globular clusters in DragonFly 44 by \cite{Saifollahi21}.

\begin{center}
\begin{figure}[H]
\includegraphics[width=1.\textwidth]{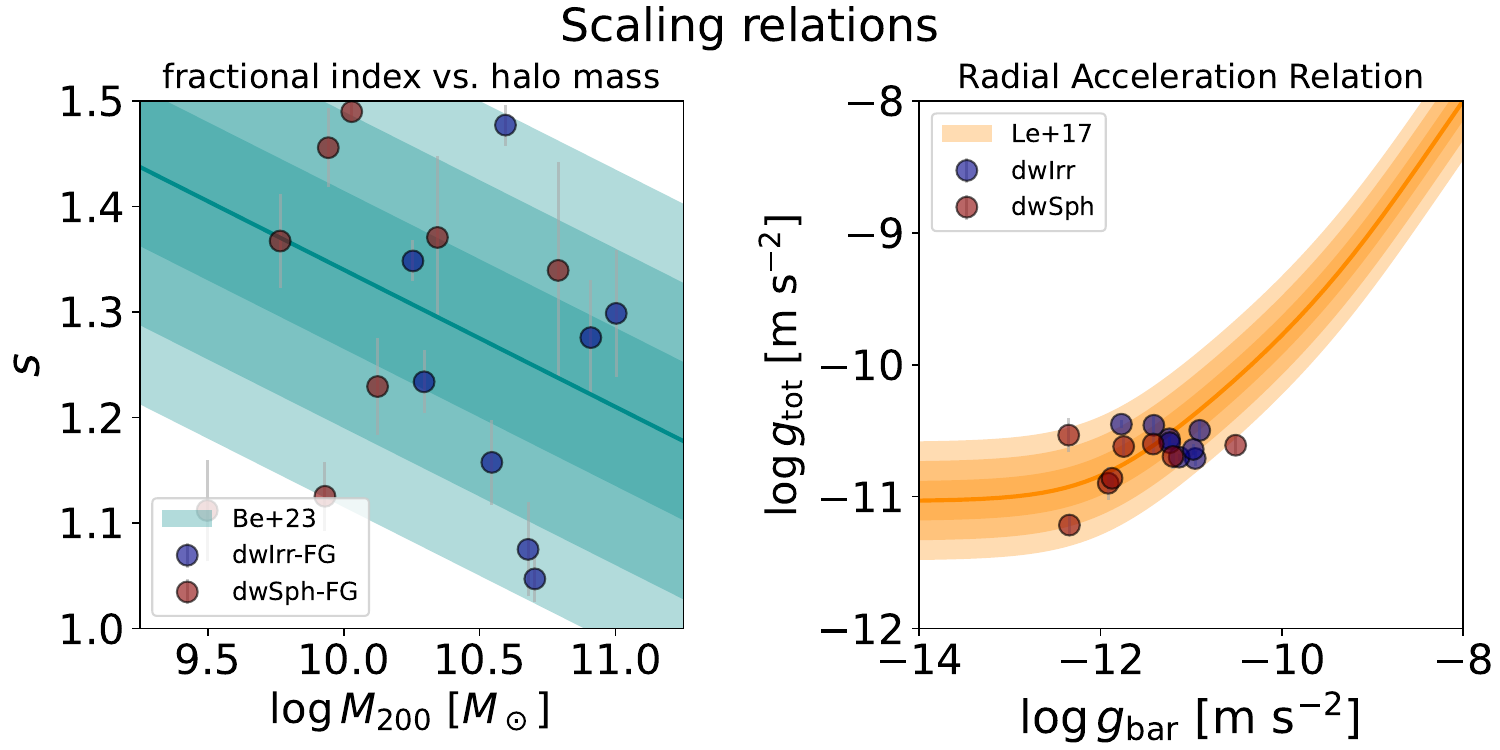}
\caption{Left: fractional index as a function of DM mass from the outcomes of our Bayesian analysis. Blue circles refer to dwIrr and red ones to dwSph; the cyan line and shaded areas illustrate the bestfit relation and $1-2-3\sigma$ dispersion from the analysis of stacked rotation curves for disc-dominated galaxies by \cite{Benetti23a}. Right: the Radial Acceleration Relation. Blue symbols refer to dwIrr and red ones to dwSph; the orange line and shaded areas illustrate the bestfit relation and $1-2-3\sigma$ dispersion from the determination by \cite{Lelli17}.}\label{fig|scaling}
\end{figure}
\end{center}

By inspecting Tables \ref{tab|results_dwIrr} and \ref{tab|results_dwSph} overall one can conclude that the evidence in favor of FG is less compelling in dwSph with respect to dwIrr. This is due to several reasons. First, the main observable for dwIrr is the rotation velocity, which is a direct probe of the mass profile; contrariwise, in dwSph the l.o.s. dispersion profile encases the mass profile in an integrated way, weighted by a kernel that depends on the tracer profiles and on the anisotropy parameter. In addition, the priors on the stellar mass-to-light ratio from population synthesis model are looser for dwSph than for dwIrr (especially when considering $3.6\,\mu$m luminosities for the latter). Finally, the observed l.o.s. dispersion profiles are more scattered and featureless with respect to the rotation curves. Thus it should not be surprising that the constraints from dwSph are less statistically significant. Nevertheless, these systems may offer an environment where any evidence in favor of FG is more robust, since the lack of baryons even in the innermost regions does not allow us to rely on different interpretations related to baryonic-induced modification of the DM profile. 

Finally, in Fig. \ref{fig|scaling} we illustrate two interesting scaling relations, that constitute relevant crosschecks of our results. The diagram on the left panel displays the fractional index $s$ as a function of the DM mass $M_{200}$. Apart for a few exceptions (objects with $s\lesssim 1.1$), the values from our analysis of individual dwarf galaxies in FG are consistent to within $2-3\sigma$ with the expectation from the relationship by \cite{Benetti23a}, that has been derived from fitting stacked rotation curves of rotation-dominated galaxies.

The diagram on the right reports the Radial Acceleration Relation or RAR \cite{McGaugh16,Lelli17} between the total acceleration $g_{\rm tot}$ and the baryonic one $g_{\rm bar}$. This is an empirical relationship known to hold for different kind of galaxies, whose average and $1-2-3\sigma$ dispersion is plotted as an orange line surrounded by shaded areas. For rotationally supported systems we compute $g_{\rm bar}=[v_{\rm gas}^2(r_{1/2})+(M_\star/L)\times v_{\rm disk}^2(r_{1/2})]/r_{1/2}$ and $g_{\rm tot}=v_{\rm tot}^2(r_{1/2})/r_{1/2}$ in terms of Equation (\ref{eq|vrot}). For dispersion-dominated systems we instead estimate $g_{\rm bar}=G\, (M_\star/L)\times L_\star/2\,r_{1/2}^2$ and $g_{\rm tot}=3\, \sigma^2_r(r_{1/2})/r_{1/2}$ in terms of Equation (\ref{eq|sigmar}). For the sake of simplicity we compute the accelerations at $r_{1/2}$, by using the bestfit values of $M_\star/L$, $M_{200}$ and $s$ from our analysis in FG. Reassuringly, almost all our estimated accelerations are consistent within $2-3\sigma$ with the RAR  by \cite{Lelli17}, with the dwIrr galaxies clustering around the value of $g_{\rm bar}$ where the relation starts to flatten, and with some dwSph tracing the flat portion of the RAR and its scatter.


\section{Summary}\label{sec|Summary}

Dark matter (DM) in fractional gravity (FG) constitutes a framework that
strikes an intermediate course between a modified gravity theory and an exotic DM scenario.
It envisages the DM component in virialized cosmic structures to be affected by a non-local interaction mediated by gravity. Specifically, in such a framework the gravitational potential associated to a given DM density distribution is determined by a modified Poisson equation including fractional derivatives, that are aimed at describing non-locality. 

Remarkably, FG can be reformulated in terms of the standard Poisson equation, but with an effective density
distribution which is flatter in the inner region with respect to the true one. Therefore FG offers a straightforward solution to the core-cusp problem of the standard $\Lambda$CDM model without altering the NFW density profile indicated by $N-$body simulations. An observer trying to interpret the kinematic data (e.g., rotation curves in dwIrr) in terms of the canonical (instead of the fractional) Poisson equation would claim the need for a cored density distribution. However, this is only apparent, since in FG the cuspy NFW density profile of $\Lambda$CDM originates a dynamics very similar to a cored profile in the standard Newtonian setting. In previous works \cite{Benetti23a,Benetti23b} we tested our FG framework by exploiting stacked rotation curves of galaxies with different masses and joint X-ray/Sunyaev-Zel’dovich observations of galaxy clusters; our analysis highlighted that the strengths of FG effects tend to weaken toward more massive systems, so implying that dwarf galaxies constitute the best environment to constrain such a scenario. 

In this paper we have dug deeper to probe FG via high-quality data of individual dwarf galaxies, by exploiting the rotation velocity profiles inferred from stellar and gas kinematic measurements in 8 dwarf irregulars, and the projected velocity dispersion profiles inferred from the observed dynamics of stellar tracers in 7 dwarf spheroidals and in the ultra-diffuse galaxy DragonFly 44.  We have found that FG reproduces extremely well the rotation and dispersion curves of all the analysed galaxies, performing in most instances significantly better than the standard Newtonian gravity. With respect to the latter, the FG fits imply slightly larger stellar mass-to-light ratios $M_\star/L$ (more in line with the values expected from galaxy colors and stellar population synthesis models), appreciably larger DM masses $M_{200}$ of a factor a few, and (for dispersion-dominated systems) more isotropic orbits. We have stressed that our bestfit determinations of the fractional index $s$ and of the DM masses $M_{200}$ from the kinematics of individual dwarf galaxies are consistent to within $2-3\sigma$ with the relationship by \cite{Benetti23a}, that has been derived from fitting stacked rotation curves of rotation-dominated galaxies. We have also highlighted that our findings are consistent with the Radial Acceleration Relation by \cite{Lelli17}.


We have pointed out that the evidence in favor of FG is less compelling in dwSph with respect to dwIrr; this is because various reasons: the l.o.s. velocity dispersion is less sensitive than the rotation velocity to the mass profile; uncertainties on the stellar mass-to-light ratio from stellar models (used as priors) are larger for dwSph than for dwIrr; the uncertainty in dispersion profile measurements are larger than in the rotation curve data. However, it should be considered that dwSph and ultra-diffuse galaxies could potentially provide a more robust environment to test FG, since they are strongly DM dominated also in the innermost regions, and thus should not have suffered from baryonic feedback processes or baryon-induced modification of the density profile. Future observations by astrometric space mission aimed at precision determination of the dispersion profiles in dwSph and ultra-diffuse galaxies will be extremely helpful to robustly strengthen the constraints on the FG framework presented here.

This work concludes a series of papers aimed at testing FG on different astrophysical scales, from dwarf galaxies to galaxy clusters. All in all, these have demonstrated that the FG framework works can solve the small scales issues of the standard $\Lambda$CDM, by reconciling with data the DM density distribution expected from $N-$body simulations, and saving its successes on large cosmological scales. Our future efforts will be directed to explain the physical origin of the nonlocal effects subtended by the FG framework, and investigate to what extent the theory can be generalized in a fully relativistic setting.

\appendix

\section*{Appendix: projection integrals}

In this Appendix we recall some useful formulas for the computation of the projection integrals involved in the analysis of Sect. \ref{sec|Methods}. We start by considering the line-of-sight velocity dispersion of Equation (\ref{eq|sigmalos}):
\begin{equation}
\sigma_{\rm los}^2(R) = \cfrac{2}{\Sigma_\star(R)}\, \int_R^\infty{\rm d}r\,\left[1-\beta\, \cfrac{R^2}{r^2}\right]\, \cfrac{\rho_\star(r)\,\sigma_r^2(r)\, r}{\sqrt{r^2-R^2}} 
\end{equation}
Numerically computing the involved integral is a bit tricky since the integrand has a singularity in the lower limit of integration. To circumvent the issue one can
insert the explicit expression for $\sigma_r$ from Equation (\ref{eq|sigmar}) and obtain
\begin{equation}
\sigma_{\rm los}^2(R) = \cfrac{2}{\Sigma_\star(R)}\, \int_R^\infty{\rm d}r\,\left[1-\beta\, \cfrac{R^2}{r^2}\right]\, \cfrac{r^{1-2\beta}}{\sqrt{r^2-R^2}} \, \int_r^\infty{\rm d}s\, s^{2\beta-1}\, \rho_\star(s)\,v_{\rm rot}^2(s)
\end{equation}
where $v_{\rm rot}^2(r)\equiv G\, [M_{\rm F}(<r)+M_\star(<r)]/r$.
It is now convenient to invert the order of integration, so that the double integral turns into
\begin{equation}
\sigma_{\rm los}^2(R) = \cfrac{2}{\Sigma_\star(R)}\,\int_R^\infty{\rm d}s\, s^{2\beta-1}\, \rho_\star(s)\,v_{\rm rot}^2(s)\,\int_R^s{\rm d}r\, 
\left[1-\beta\, \cfrac{R^2}{r^2}\right]\, \cfrac{r^{1-2\beta}}{\sqrt{r^2-R^2}} \, 
\end{equation}
and the inner integral can be expressed in terms of special functions.

The overall results can be written as \cite{Mamon05}
\begin{equation}
\sigma_{\rm los}^2(R) = \cfrac{2}{\Sigma_\star(R)}\,\int_R^\infty{\rm d}s\,  \rho_\star(s)\,v_{\rm rot}^2(s)\, \mathcal{K}_2(s/R,\beta)
\end{equation}
in terms of the kernel
\begin{equation}\label{eq|kerneò}
\begin{aligned}
\mathcal{K}_2(x,\beta) &= x^{2\beta-1}\,\left[-\cfrac{1}{2}\,B\left(\beta-\cfrac{1}{2},\cfrac{1}{2};\cfrac{1}{x^2}\right)+\cfrac{\beta}{2}\,B\left(\beta+\cfrac{1}{2},\cfrac{1}{2};\cfrac{1}{x^2}\right)\right.+\\
&\\
&\left.+ \cfrac{3-2\beta}{4}\,\cfrac{\sqrt{\pi}\,\Gamma(\beta-1/2)}{\Gamma(\beta)}\right]~,
\end{aligned}
\end{equation}
where $\Gamma(a) = \int_0^\infty\,{\rm d}t\, t^{a-1}\, e^{-t}$ is the Gamma function and $B(a,b;x)\equiv \int_0^x{\rm d}t\, t^{a-1}\, (1-t)^{b-1}$ is the incomplete Beta function.

\vspace{6pt}

\authorcontributions{Conceptualization: F.B., A.L., G.G.; methodology: F.B., A.L., G.G.; validation: M.A.B., Y.B., B.S.H., C.B.; writing: A.L. All authors have read and agreed to the published
version of the manuscript.}

\funding{This work was partially funded from the projects: ``Data Science methods for MultiMessenger Astrophysics \& Multi-Survey Cosmology'' funded by the Italian Ministry of University and Research, Programmazione triennale 2021/2023 (DM n.2503 dd. 9 December 2019), Programma Congiunto Scuole; EU H2020-MSCA-ITN-2019 n. 860744 \textit{BiD4BESt: Big Data applications for black hole Evolution STudies}; PRIN MIUR 2017 prot. 20173ML3WW, \textit{Opening the ALMA window on the cosmic evolution of gas, stars, and supermassive black holes}; Fondazione ICSC, Spoke 3 Astrophysics and Cosmos Observations; National Recovery and Resilience Plan (Piano Nazionale di Ripresa e Resilienza, PNRR) Project ID CN-00000013 ``Italian Research Center on High-Performance Computing, Big Data and Quantum Computing'' funded by MUR Missione 4 Componente 2 Investimento 1.4: Potenziamento strutture di ricerca e creazione di ``campioni nazionali di R\&S (M4C2-19)''---Next Generation EU (NGEU); INAF Large Grant 2022 funding scheme with the project ``MeerKAT and LOFAR Team up: a Unique Radio Window on Galaxy/AGN co-Evolution''.}

\dataavailability{N/A}

\acknowledgments{We thank the anonymous referees for useful comments and suggestions. We acknowledge L. Danese, S. Liberati and P. Salucci for helpful discussions.}

\conflictsofinterest{The authors declare no conflict of interest.}

\begin{adjustwidth}{-\extralength}{0cm}

\reftitle{References}
\printendnotes[custom]

\end{adjustwidth}


\begin{thebibliography}{999}

\bibitem{Rubin80}Rubin, V.C.; Ford, W.K.,Jr.; Thonnard, N.  {Rotational properties of $21$ Sc galaxies with a large range of luminosities and radii, from NGC 4605 ($R = 4$ kpc) to UGC 2885 ($R=122$ kpc)}. \emph{Astrophys. J.} \textbf{1996}, \emph{281}, 27.

\bibitem{Persic96}Persic, M.; Salucci, P.; Stel, F. {The universal rotation curve of spiral galaxies — I. The dark matter connection}. \emph{Mon. Not. R. Astron. Soc.} \textbf{1980}, \emph{238}, 471.

\bibitem{Navarro97}Navarro, J.F.; Frenk, C.S.; White, S.D.M. {A Universal Density Profile from Hierarchical Clustering}. \emph{Astrophys. J.} \textbf{1997}, \emph{490}, 493.

\bibitem{Dutton14}Dutton, A.A.; Maccio, A.V. {Cold dark matter haloes in the Planck era: Evolution of structural parameters for Einasto and NFW profiles}.  \emph{Mon. Not. R. Astron. Soc.} \textbf{2014}, \emph{441}, 3359..

\bibitem{Wang20}Wang, J.; Bose, S.; Frenk, C. S.; Gao, L.; Jenkins, A.; Springel, V.; White, S. D. M. {Universal structure of dark matter haloes over a mass range of 20 orders of magnitude}. \emph{Nature} \textbf{2020}, \emph{585}, 39.

\bibitem{Flores94}Flores, R.A.; Primack, J.R. {Observational and Theoretical Constraints on Singular Dark Matter Halos}. \emph{Astrophys. J.} \textbf{1994}, \emph{427}, L1.

\bibitem{Gentile04}Gentile, G.; Salucci, P.; Klein, U.; Vergani, D.; Kalberla, P. {The cored distribution of dark matter in spiral galaxies}. \emph{Mon. Not. R. Astron. Soc.} \textbf{2004}, \emph{351}, 903.

\bibitem{deBlok08}de Blok, W. J. G.; Walter, F.; Brinks, E.; Trachternach, C.; Oh, S. -H.; Kennicutt, R. C., Jr.
{High-Resolution Rotation Curves and Galaxy Mass Models from THINGS}. \emph{Astron. J.} \textbf{2008}, \emph{136}, 2648.

\bibitem{Oh15}Oh, S.-H.; Hunter, D. A.; Brinks, E.; Elmegreen, B.G.; Schruba, A.; Walter, F.; Rupen, M.P.; Young, L.M.; Simpson, C.E.; Johnson, M.C.; et al.  
{High-resolution Mass Models of Dwarf Galaxies from LITTLE THINGS}. \emph{Astron. J.} \textbf{2015}, \emph{149}, 1800.

\bibitem{Bullock17}Bullock, J.S.; Boylan-Kolchin, M.
{Small-Scale Challenges to the $\Lambda$CDM Paradigm}. \emph{Ann. Rev. Astron. Astrophys.} \textbf{2017}, \emph{55}, 343.

\bibitem{Klypin99}Klypin, A.; Kravtsov, A.V.; Valenzuela, O.; Prada, F. {Where Are the Missing Galactic Satellites?}. \emph{Astrophys. J.} \textbf{1999}, \emph{522}, 82.

\bibitem{Moore99}Moore, B.; Ghigna, S.; Governato, F.; Lake, G.; Quinn, T.; Stadel, J.; Tozzi, P. {Dark Matter Substructure within Galactic Halos}. \emph{Astrophys. J.} \textbf{1999}, \emph{524}, L19.

\bibitem{BoylanKolchin12}Boylan-Kolchin, M.; Bullock, J.S.; Kaplinghat, M. {The Milky Way's bright satellites as an apparent failure of $\Lambda$CDM}. \emph{Mon. Not. R. Astron. Soc.} \textbf{2012}, \emph{422}, 1203.

\bibitem{McGaugh16}McGaugh, S.S.; Lelli, F.; Schombert, J.M. {Radial Acceleration Relation in Rotationally Supported Galaxies}. \emph{Phys. Rev. Lett.} \textbf{2016}, \emph{117}, 1101.

\bibitem{Lelli17}Lelli, F.; McGaugh, S.S.; Schombert, J.M.; Pawlowski, M.S. {One Law to Rule Them All: The Radial Acceleration Relation of Galaxies}. \emph{Astrophys. J.} \textbf{2017}, \emph{836}, 152.

\bibitem{Donato09}Donato, F.; Gentile, G.; Salucci, P.; Frigerio Martins, C.; Wilkinson, M.I.; Gilmore, G.; Grebel, E.K.; Koch, A.; Wyse, R. {A constant dark matter halo surface density in galaxies}. \emph{Mon. Not. R. Astron. Soc.} \textbf{2009}, \emph{397}, 1169.

\bibitem{Donato04}Donato, F.; Gentile, G.; Salucci, P. {Cores of dark matter haloes correlate with stellar scalelengths}. \emph{Mon. Not. R. Astron. Soc.} \textbf{2009}, \emph{353}, L17.

\bibitem{Oman15}Oman, K.A.; Navarro, J.F.; Fattahi, A.; Frenk, C.S.; Sawala, T.; White, S.D.M.; Bower, R.; Crain, R.A.; Furlong, M.; Schaller, M.; et al. {The unexpected diversity of dwarf galaxy rotation curves}. \emph{Mon. Not. R. Astron. Soc.} \textbf{2009}, \emph{452}, 3650.

\bibitem{Pontzen14}Pontzen, A.; Governato, F. {Cold dark matter heats up}. \emph{Nature} \textbf{2014}, \emph{506}, 171.

\bibitem{Freundlich20}Freundlich, J.; Dekel, A.; Jiang, F.; Ishai, G. ; Cornuault, N.; Lapiner, S.; Dutton, A.A.; Maccio, A.V.  {A model for core formation in dark matter haloes and ultra-diffuse galaxies by outflow episodes}. \emph{Mon. Not. R. Astron. Soc.} \textbf{2020}, \emph{491}, 4253.

\bibitem{ElZant01}El-Zant, A.; Shlosman, I.; Hoffman, Y.  {Dark Halos: The Flattening of the Density Cusp by Dynamical Friction}. \emph{Astrophys. J.} \textbf{2001}, \emph{560}, 636.

\bibitem{Tonini06}Tonini, C., Lapi, A., Salucci, P. {Angular Momentum Transfer in Dark Matter Halos: Erasing the Cusp}. \emph{Astrophys. J.} \textbf{2006}, \emph{649}, 591.

\bibitem{Bode01}Bode, P.; Ostriker, J.P.; Turok, N. {Halo Formation in Warm Dark Matter Models}. \emph{Astrophys. J.} \textbf{2001}, \emph{556}, 93.

\bibitem{Lovell14}Lovell, M.; Frenk, C.S.; Eke, V.R.; Jenkins, A.; Gao, L.; Theuns, T. {The properties of warm dark matter haloes}. \emph{Mon. Not. R. Astron. Soc.} \textbf{2014}, \emph{439}, 300.

\bibitem{Hu00}Hu, W.; Barkana, R.; Gruzinov, A. {Fuzzy Cold Dark Matter: The Wave Properties of Ultralight Particles}. \emph{Phys. Rev. Lett.} \textbf{2000}, \emph{85}, 1158.

\bibitem{Hui17}Hui, L.; Ostriker, J.P.; Tremaine, S.; Witten, E. {Ultralight scalars as cosmological dark matter}. \emph{Phys. Rev. D} \textbf{2017}, \emph{95}, 3541.

\bibitem{Vogelsberger16}Vogelsberger, M.; Zavala, J.; Cyr-Racine, F.-Y.; Pfrommer, C.; Bringmann, T.; Sigurdson, K. {ETHOS-An effective theory of structure formation: Dark matter physics as a possible explanation of the small-scale CDM problems}. \emph{Mon. Not. R. Astron. Soc.} \textbf{2016}, \emph{460}, 1399.

\bibitem{McDermott20}McDermott, S.D.; Witte, S.J.
{Cosmological evolution of light dark photon dark matter}. \emph{Phys. Rev. D} \textbf{2020}, \emph{101}, 3030.

\bibitem{Bolton22}Bolton, J.S.; Caputo, A.; Liu, H.; Viel, M.
{Comparison of Low-Redshift Lyman-$\alpha$ Forest Observations to Hydrodynamical Simulations with Dark Photon Dark Matter}. \emph{Phys. Rev. Lett.} \textbf{2022}, \emph{129}, 1102.

\bibitem{Bertone04}Bertone, G.; Hooper, D.; Silk, J. {Particle dark matter: Evidence, candidates and constraints}. \emph{Phys. Rev.} \textbf{2004}, \emph{405}, 279.

\bibitem{Adhikari17}Adhikari, R.; Agostini, M.; Ky, N.A.; Araki, T.; Archidiacono, M.; Bahr, M.; Baur, J.; Behrens, J.; Bezrukov, F.; Bhupal Dev, P. S.; et al.  {A White Paper on keV sterile neutrino Dark Matter}. \emph{Journ. Cosmol. Astropart. Phys.} \textbf{2017}, \emph{1}, 25.

\bibitem{Salucci21}Salucci, P.; Esposito, G.; Lambiase, G.; Battista, E.; Benetti, M.; Bini, D.; Boco, L.; Sharma, G.; Bozza, V.; Buoninfante, L.; et al. {Einstein, Planck and Vera Rubin: Relevant encounters between the Cosmological and the Quantum Worlds}. \emph{Front. Phys}. \textbf{2021}, \emph{8},~603190.

\bibitem{Clifton12}Clifton, T.; Ferreira, P.G.; Padilla, A,; Skordis, C. {Modified gravity and cosmology.} \emph{Phys. Rep.} \textbf{2012}, \emph{513}, 1.

\bibitem{Nojiri17}Nojiri, S.; Odintsov, S.D.; Oikonomou, V.K. {Modified gravity theories on a nutshell: Inflation, bounce and late-time evolution}. \emph{Phys. Rep.} \textbf{2017}, \emph{692}, 1.

\bibitem{Saridakis21}Saridakis, E.N.; Lazkoz, R.; Salzano, V.; Moniz, P.V.; Capozziello, S; Beltran J.J.; De Laurentis, M.; Olmo, G.J.  {Modified Gravity and Cosmology: An Update by the CANTATA Network}. \emph{ISBN: 978-3-030-83715-0} \textbf{2021} { Cham: Springer International Publishing}.

\bibitem{Milgrom83}Milgrom, M. {A modification of the Newtonian dynamics as a possible alternative to the hidden mass hypothesis}. \emph{Astrophys. J.} \textbf{2017}, \emph{270}, 365.

\bibitem{Famaey12}Famaey, B.; McGaugh, S.S. {Modified Newtonian Dynamics (MOND): Observational Phenomenology and Relativistic Extensions}. \emph{Liv. Rev. Relat.} \textbf{2012}, \emph{15}, 10.

\bibitem{Varieschi20}Varieschi, G.U. {Newtonian Fractional-Dimension Gravity and MOND}. \emph{Found. Phys.} \textbf{2020}, \emph{50}, 1608.

\bibitem{Varieschi21}Varieschi, G.U. {Newtonian fractional-dimension gravity and rotationally supported galaxies}. \emph{Mon. Not. R. Astron. Soc.} \textbf{2021}, \emph{503}, 1915.

\bibitem{Giusti20a}Giusti, A. {MOND-like fractional Laplacian theory}.  \emph{Phys. Rev. D} \textbf{2020}, \emph{101}, 124029.

\bibitem{Giusti20b}Giusti, A.; Garrappa, R.; Vachon, G. {On the Kuzmin model in fractional Newtonian gravity}. \emph{EPJP} \textbf{2020}, \emph{135}, 798.

\bibitem{Calcagni22}Calcagni, G.; Varieschi, G.U. {Gravitational potential and galaxy rotation curves in multi-fractional spacetimes}. \emph{J. High Energy Phys.} \textbf{2022}, \emph{8}, 24.

\bibitem{Verlinde17}Verlinde, E.P. {Emergent Gravity and the Dark Universe}. \emph{Sci. Post. Phys.} \textbf{2017}, \emph{2}, 16.

\bibitem{Yoon23}Yoon, Y.; Park, J.-C.; Hwang, H.S. {Understanding galaxy rotation curves with Verlinde's emergent gravity.} \emph{Class. Quant. Grav.} \textbf{2023}, \emph{40}, 02LT01.

\bibitem{Benetti23a}Benetti, F.; Lapi, A.; Gandolfi, G.; Salucci, P.; Danese, L. {Dark Matter in Fractional Gravity I: Astrophysical Tests on Galactic Scales}. \emph{Astrophys. J.}
\textbf{2023}, \emph{949}, 65.

\bibitem{Benetti23b}Benetti, F.; Lapi, A.; Gandolfi, G.; Haridasu, B.S.; Danese, L. {Dark Matter in Fractional Gravity II: Tests in Galaxy Clusters}. \emph{Universe}
\textbf{2023}, \emph{9}, 329.

\bibitem{Gandolfi21}Gandolfi, G.; Lapi, A.; Liberati, S. {Self-gravitating Equilibria of Non-minimally Coupled Dark Matter Halos}. \emph{Astrophys. J.} \textbf{2021}, \emph{910}, 76.

\bibitem{Gandolfi22}Gandolfi, G.; Lapi, A.; Liberati, S. {Empirical Evidence of Nonminimally Coupled Dark Matter in the Dynamics of Local Spiral Galaxies?} \emph{Astrophys. J.} \textbf{2022}, \emph{929}, 48.

\bibitem{Gandolfi23}Gandolfi, G.; Haridasu, B.S.; Liberati, S.; Lapi, A. {Looking for Traces of Non-minimally Coupled Dark Matter in the X-COP Galaxy Clusters Sample}. \emph{Astrophys. J.} \textbf{2023}, \emph{952}, 105.

\bibitem{Aghanim20}Aghanim, N.; Akrami, Y.; Ashdown, M.; Aumont, J.; Baccigalupi, C.; Ballardini, M.; Banday, A. J.; Barreiro, R. B.; Bartolo, N.; Basak, S.; et al. [Planck Collaboration]. {Planck 2018 results. VI. Cosmological parameters}. \emph{Astron. Astrophys.} \textbf{2020}, \emph{641}, A6.

\bibitem{Uchaikin13}Uchaikin, V.V. {Fractional Derivatives for Physicists and Engineers: Background and Theory}. \emph{Nonlinear Physical Science, Higher Education Press: Beijing and Springer-Verlag: Berlin Heidelberg} \textbf{2013}, ISBN 978-3-642-33910-3. 

\bibitem{GomezAguilar12}Gomez-Aguilar, J.F.; Rosales-Garcıa, J.J.; Bernal-Alvarado, J.J.; Cordova-Fraga, T; Guzman-Cabrera, R.  {Fractional mechanical oscillators}. \emph{Riv. Mex. Phys.}
\textbf{2012}, \emph{58}, 348.

\bibitem{Ebaid19}Ebaid, A.; El-Zahar, E.R.; Aljohani, A.F.;
Salah, B.; Krid, M.; Tenreiro Machado, J. {Analysis of the two-dimensional fractional projectile motion
in view of the experimental data}. \emph{Nonlin. Dynam.}
\textbf{2019}, \emph{97}, 1711.

\bibitem{Pavan19}Pranjivan Mehta, P.; Pang, G.; Song, F.;
Em Karniadakis, G. {Discovering a Universal Variable-Order Fractional Model for Turbulent Couette Flow Using a Physics-informed Neural Network}. \emph{Fract. Calc. Appl. Analys.}
\textbf{2019}, \emph{22}, 6.

\bibitem{Binney82}Binney, J. {Dynamics of elliptical galaxies and other spheroidal components}. \emph{Ann. Rev. Astron. Astrophys.} \textbf{1982}, \emph{20}, 399.

\bibitem{Lokas01}Lokas, E.L.; Mamon, G.A. {Properties of spherical galaxies and clusters with an NFW density profiles}. \emph{Mon. Not. R. Astron. Soc.} \textbf{2001}, \emph{321}, 155.

\bibitem{Plummer11}Plummer, H. C.
{On the problem of distribution in globular star clusters}. \emph{Mon. Not. R. Astron. Soc.} \textbf{2001}, \emph{71}, 460.

\bibitem{Lelli16}Lelli, F.; McGaugh, S.S.; Schombert, J.M. {SPARC: Mass Models for 175 Disk Galaxies with Spitzer Photometry and Accurate Rotation Curves}. \emph{Astron. J.} \textbf{2016}, \emph{152}, 157.

\bibitem{Li20}Li, P.; Lelli, F.; McGaugh, S.; Schombert, J. {A Comprehensive Catalog of Dark Matter Halo Models for SPARC Galaxies}. \emph{Astrophys. J. Suppl. Ser.} \textbf{2020}, \emph{247}, 31.

\bibitem{Walker09}Walker, M.G.; Mateo, M.; Olszewski, E.W.  {Stellar Velocities in the Carina, Fornax, Sculptor, and Sextans dSph Galaxies: Data From the Magellan/MMFS Survey}. \emph{Astron. J.} \textbf{2009}, \emph{137}, 3100.

\bibitem{Mateo08}Mateo, M.; Olszewski, E.W.; Walker, M.G.  {The Velocity Dispersion Profile of the Remote Dwarf Spheroidal Galaxy Leo I: A Tidal Hit and Run?}. \emph{Astrophys. J.} \textbf{2009}, \emph{675}, 201.

\bibitem{DeMartino23}de Martino, I.; Diaferio, A.; Ostorero, L.
{Dynamics of dwarf galaxies in $f(R)$ gravity}. \emph{Mon. Not. R. Astron. Soc.} \textbf{2023}, \emph{519}, 4424.

\bibitem{vanDokkum19}van Dokkum, P.; Wasserman, A.; Danieli, S.; Abraham, R.; Brodie, J.; Conroy, C.; Forbes, D.A.; Martin, C.; Matuszewski, M.; Romanowsky, A.J.; Villaume, A. 
{Spatially Resolved Stellar Kinematics of the Ultra-diffuse Galaxy Dragonfly 44. I. Observations, Kinematics, and Cold Dark Matter Halo Fits}. \emph{Astrophys. J.} \textbf{2019}, \emph{880}, 91.

\bibitem{Leisman17}Leisman, L.; Haynes, M.P.; Janowiecki, S.; Hallenbeck, G.; Jozsa, G.; Giovanelli, R.; Adams, E.A.K.; Bernal Neira, D.; Cannon, J.M.; Janesh, W.F.; et al. {(Almost) Dark Galaxies in the ALFALFA Survey: Isolated H I-bearing Ultra-diffuse Galaxies}. \emph{Astrophys. J.}  \textbf{2017}, \emph{842}, 133.

\bibitem{ManceraPina20}Mancera Pina, P.E.; Fraternali, F.; Oman, K.A.; Adams, E.A.K.; Bacchini, C.; Marasco, A.; Oosterloo, T.; Pezzulli, G.; Posti, L.; et al. {Robust H I kinematics of gas-rich ultra-diffuse galaxies: hints of a weak-feedback formation scenario}. \emph{Mon. Not. R. Astron. Soc.}  \textbf{2020}, \emph{495}, 3636.

\bibitem{ManceraPina19}Mancera Piña, P.E.; Aguerri, J.A.L.; Peletier, R.F.; Venhola, A.; Trager, S.; Choque Challapa, N. {The evolution of ultra-diffuse galaxies in nearby galaxy clusters from the Kapteyn IAC WEAVE INT Clusters Survey}. \emph{Mon. Not. R. Astron. Soc.}  \textbf{2019}, \emph{485}, 1036.

\bibitem{Sengupta19}Sengupta, C.; Scott, T. C.; Chung, A.; Wong, O. I. {Dark matter and H I in ultra-diffuse galaxy UGC 2162}. \emph{Mon. Not. R. Astron. Soc.}  \textbf{2019}, \emph{488}, 3222.

\bibitem{vanDokkum15}van Dokkum, P.G.; Abraham, R.; Merritt, A.; Zhang, J.; Geha, M.; Conroy, C. {Forty-seven Milky Way-sized, Extremely Diffuse Galaxies in the Coma Cluster}. \emph{Astrophys. J.}  \textbf{2015}, \emph{798}, L45.

\bibitem{Wasserman19}Wasserman, A.; van Dokkum, P.; Romanowsky, A.J.; Brodie, J.; Danieli, S.; Forbes, D.A.; Abraham, R.; Martin, C.; Matuszewski, M.; Villaume, A.; Tamanas, J.; Profumo, S. {Spatially Resolved Stellar Kinematics of the Ultra-diffuse Galaxy Dragonfly 44. II. Constraints on Fuzzy Dark Matter}. \emph{Astrophys. J.} \textbf{2019}, \emph{885}, 155.

\bibitem{Julio23}Julio, M.P.; Brinchmann, J.; Zoutendijk, S.L.; Read, J.I.; Vaz, D.; Kamann, S.; Krajnovic, D.; Boogaard, L.A.; Steinmetz, M.; Bouche, N. {The MUSE-Faint survey. V. Constraining Scalar Field Dark Matter with Antlia B}. \emph{Astron. and Astrophys.} \textbf{2023}, 678, A38

\bibitem{Bell01}Bell, E.F.; de Jong, R.S.
{Stellar Mass-to-Light Ratios and the Tully-Fisher Relation}. \emph{Astrophys. J.} \textbf{2001}, \emph{550}, 212.

\bibitem{Portinari04}Portinari, L. ; Sommer-Larsen, J. ; Tantalo, R.
{On the mass-to-light ratio and the initial mass function in disc galaxies}. \emph{Mon. Not. R. Astron. Soc.} \textbf{2004}, \emph{346},691.

\bibitem{Schombert19}Schombert, J. McGaugh, S. Lelli, F.
{The mass-to-light ratios and the star formation histories of disc galaxies}. \emph{Mon. Not. R. Astron. Soc.} \textbf{2019}, \emph{483}, 1496.

\bibitem{Moster13}Moster, B.P.; Naab, T.; White, S.D.M.
{Galactic star formation and accretion histories from matching galaxies to dark matter haloes}. \emph{Mon. Not. R. Astron. Soc.} \textbf{2013}, \emph{428}, 3121.

\bibitem{Lange19}Lange, J.U.; van den Bosch, F.C.; Zentner, A.R.; Wang, K.; Villarreal, A.S.
{Updated results on the galaxy-halo connection from satellite kinematics in SDSS}. \emph{Mon. Not. R. Astron. Soc.} \textbf{2019}, \emph{487}, 3112.

\bibitem{Lapi18}Lapi, A.; Salucci, P. ; Danese, L.
{Precision Scaling Relations for Disk Galaxies in the Local Universe}. \emph{Astrophys. J.} \textbf{2018}, \emph{859}, 2.

\bibitem{Mandelbaum16}Mandelbaum, R.; Wang, W.; Zu, Y.; White, S.; Henriques, B.; More, S.
{Strong bimodality in the host halo mass of central galaxies from galaxy-galaxy lensing}. \emph{Mon. Not. R. Astron. Soc.} \textbf{2016}, \emph{457}, 3002.

\bibitem{ForemanMackey13}Foreman-Mackey, D.; Hogg, D.W.; Lang, D.; Goodman, J. {emcee: The MCMC Hammer}. \emph{Publ. Astron. Soc. Pac.} \textbf{2013}, \emph{125}, 306.

\bibitem{Swaters02}Swaters, R. A.; van Albada, T. S.; van der Hulst, J. M.; Sancisi, R. {The Westerbork HI survey of spiral and irregular galaxies. I. HI imaging of late-type dwarf galaxies}. \emph{Astron. Astrophys.} \textbf{2002}, \emph{390}, 829.

\bibitem{Saifollahi21}Saifollahi, Teymoor; Trujillo, Ignacio; Beasley, Michael A.; Peletier, Reynier F.; Knapen, Johan H.  {The number of globular clusters around the iconic UDG DF44 is as expected for dwarf galaxies}. \emph{Mon. Not. R. Astron. Soc.}  \textbf{2019}, \emph{502}, 5921.

\bibitem{Mamon05}Mamon, Gary A.; Lokas, E. L. {Dark matter in elliptical galaxies - II. Estimating the mass within the virial radius}. \emph{Mon. Not. R. Astron. Soc.} \textbf{2016}, \emph{363}, 705.


\end{thebibliography}
\end{document}